\documentclass[trackchanges,twocolumn]{aastex701}
\usepackage{amsmath}
\usepackage{amssymb}
\usepackage{tablefootnote}

\begin{document}

\title{JWST Imaging of Edge-on Protoplanetary Disks. V. Flux Ratio Reversal and Disk-Edge Structure in the Flying Saucer }


\author[orcid=0000-0003-3583-6652]{Jie Ma}
\altaffiliation{Swiss PostDoc.Mobility Fellow}
\affiliation{Univ. Grenoble Alpes, CNRS, IPAG, F-38000 Grenoble, France}
\email[show]{jie.ma@univ-grenoble-alpes.fr}  

\author[orcid=0000-0002-2805-7338]{Karl R. Stapelfeldt} 
\affiliation{Jet Propulsion Laboratory, California Institute of Technology, 4800 Oak Grove Drive, Pasadena, CA 91109, USA}
\email{karl.r.stapelfeldt@jpl.nasa.gov}

\author[orcid=0000-0002-8962-448X]{Marion Villenave} 
\affiliation{Univ. Grenoble Alpes, CNRS, IPAG, F-38000 Grenoble, France}
\email{marion.villenave@univ-grenoble-alpes.fr}

\author[orcid=0000-0002-5092-6464]{Gaspard Duchêne} 
\affiliation{Univ. Grenoble Alpes, CNRS, IPAG, F-38000 Grenoble, France}
\affiliation{Department of Astronomy, University of California, Berkeley CA, 94720, USA}
\email{gaspard.duchene@univ-grenoble-alpes.fr}

\author[orcid=0000-0002-1637-7393]{François Ménard} 
\affiliation{Univ. Grenoble Alpes, CNRS, IPAG, F-38000 Grenoble, France}
\email{francois.menard@univ-grenoble-alpes.fr}

\author[orcid=0000-0003-1451-6836]{Ryo Tazaki} 
\affiliation{Department of Earth Science and Astronomy, The University of Tokyo, Tokyo 153-8902, Japan}
\email{rtazaki@g.ecc.u-tokyo.ac.jp}

\author[orcid=0000-0003-2640-5586]{Youcef I. Bouarour}
\affiliation{Univ. Grenoble Alpes, CNRS, IPAG, F-38000 Grenoble, France}
\email{youcef-islam.bouarour@univ-grenoble-alpes.fr}

\author[orcid=0000-0002-7002-8928]{Greta Guidi} 
\affiliation{Univ. Grenoble Alpes, CNRS, IPAG, F-38000 Grenoble, France}
\email{greta.guidi@univ-grenoble-alpes.fr}

\author[orcid=0009-0001-8094-4735]{Laurine Martinien} 
\affiliation{Univ. Grenoble Alpes, CNRS, IPAG, F-38000 Grenoble, France}
\email{laurine.martinien@univ-grenoble-alpes.fr}

\author[orcid=0009-0003-4758-4285]{Maxime Roumesy} 
\affiliation{Univ. Grenoble Alpes, CNRS, IPAG, F-38000 Grenoble, France}
\email{maxime.roumesy@univ-grenoble-alpes.fr}

\author[orcid=0000-0001-5907-5179]{Christophe Pinte}
\affiliation{Univ. Grenoble Alpes, CNRS, IPAG, F-38000 Grenoble, France}
\email{christophe.pinte@univ-grenoble-alpes.fr}




\begin{abstract}

We present near- and mid-infrared broadband imaging observations of the edge-on protoplanetary disk 2MASS J16281370-2431391, a.k.a. the 'Flying Saucer', obtained with NIRCam and MIRI on board JWST. These data are combined with archival optical and near-IR broadband imaging from HST, millimeter dust continuum emission from ALMA, and optical lightcurves from the K2 and TESS for a comprehensive multi-wavelength study. 
The JWST observations reveal a reversal of the flux ratio between the two reflection nebulae at near-infrared wavelengths (2–3$\mu$m), which can be explained by a misaligned inner disk. For the first time, JWST reveals the bright lining along the disk edge at 2 and 3$\mu$m, arising from forward scattering of background photons by dust grains near the $\tau\approx1$ layer. By combining the dust lane thickness measured with HST and JWST, we find the dust grains up to at least $\sim$3$\mu$m are vertically mixed throughout the disk and have undergone significant growth. In addition, comparison of earlier HST/NICMOS J and K band images with those from JWST/NIRCam reveals variability in the scattered-light morphology, as well as measurable proper motion of the system with respect to the background stars, confirming that the Flying Saucer is co-moving with the young star population of $\rho$ Ophiuchi.


\end{abstract}
\keywords{\uat{Protoplanetary disks}{1300} --- \uat{Radiative transfer}{1335} --- \uat{Planet formation}{1241} --- \uat{Dust continuum emission}{412}}

\section{Introduction} \label{sec:intro}
Advances in high angular resolution imaging have revealed a wide range of protoplanetary disk morphologies across a broad wavelength range \citep{Andrews2020, Benisty2023}. While most disks are observed at moderate inclinations, edge-on disks provide a unique perspective by blocking direct starlight and presenting their vertical structure for a direct view \citep{Watson2004, Watson2007, Duchene2024, Villenave2020}. When viewed against a bright background, edge-on disks may appear in silhouette, allowing multi-wavelength disk radial and vertical thickness measurements to look for a hint of dust growth \citep{McCaughrean1998}. Therefore, such silhouette disks offer powerful diagnostics of dust evolution. 


The Flying Saucer (2MASS J16281370-2431391) is one of the best-known examples of an edge-on protoplanetary disk. It was first identified at the periphery of the $\rho$~Ophiuchi cloud in near-infrared observations by \citet{Grosso2003}. The system exhibits the characteristic pair of reflection nebulae separated by a dark lane, and has an apparent radius of 2.15\arcsec and a vertical extent of 1.2\arcsec. 
At the outset, the wavelength-dependent brightness ratio of the two reflection nebulae was noted, with a brightness reversal from the $H$ to $Ks$ bands, with the southern nebula becoming brighter at longer wavelengths \citep{Grosso2003}. 
Mid-infrared spectroscopy of the Flying Saucer with Spitzer revealed the characteristic spectral energy distribution (SED) of an edge-on optically thick disk, and the brightness of the mid-IR SED showed that micron-sized grains must be present at the disk surface \citep{Pontoppidan2007}. They also reported detections of H$_2$ from the warm gas and tentative polycyclic
aromatic hydrocarbon (PAH) emission, although the emission origin was not constrained.

At longer wavelengths, high-resolution millimeter observations have shown that the gas disk extends beyond the dust disk. The CO emission at 242 GHz reaches radii of 2.5\arcsec \citep{Guilloteau2016}, while the millimeter dust continuum is more compact, only extending to 1.6\arcsec in radius at 0.87 and 1.3mm \citep{Guilloteau2025}. Such disk extent difference is commonly interpreted as a signature of radial drift, although optical depth effects and observational limitations may also contribute to the apparent radial extent \citep{Facchini2017}.

In the CO molecular line maps, the Flying Saucer appears in silhouette against the extended background CO emission \citep{Guilloteau2016}. 
More recently, JWST observations resolved the disk in silhouette at optical and mid-infrared wavelengths, where it is seen in absorption against spatially extended background PAH and $\rm H_2$ emission \citep{Dartois2025}. The silhouette along the disk midplane extends radially beyond the Atacama Large Millimeter/submillimeter Array (ALMA) millimeter continuum disk and the scattered-light lobes above and below the midplane seen in the outer disk region, suggesting the presence of grains up to tens of microns at high altitude. 

Despite being extensively observed, several questions about the Flying Saucer remain open. In particular, the physical origin of the wavelength-dependent brightness ratio between the two nebulae and its connection to disk geometry and dust properties have not yet been studied with a unified detailed model. With JWST now opening the near- and mid-infrared regimes at high sensitivity and angular resolution, the Flying Saucer is an ideal case to study the disk structure, the scattering behavior, and dust evolution. 

In this paper, we aim to investigate the origin of the wavelength-dependent flux ratio of the two scattered-light lobes of the system, and the nature of the silhouette and newly detected lining features in the Flying Saucer.   
In Section~\ref{sec:obs}, we describe the JWST observations and the archival dataset used in this study, together with the data reduction procedures. In Section~\ref{sec:result}, we present our analysis on the disk morphology, including the detection of the lining along the disk outer edge, the first estimation of the system's proper motion, and the variation of the flux ratio of the two nebulae across multiple wavelengths and epochs. In Section~\ref{sec:model}, we construct radiative transfer disk models to explore the physical origin of the flux ratio reversal across the near-infrared, and the observed silhouette features. Then in Section~\ref{sec:discussion}, we place the Flying Saucer in a broader context of other edge-on disks and discuss the implications of the flux ratio reversal and the silhouette morphology for disk geometry and dust properties. The conclusions are summarized in Section~\ref{sec: conclusion}. 

\section{Observation and Data Reduction} \label{sec:obs}
\subsection{JWST} \label{subsec:obs-jwst}
\begin{figure}
    \centering
    \includegraphics[width=0.9\linewidth]{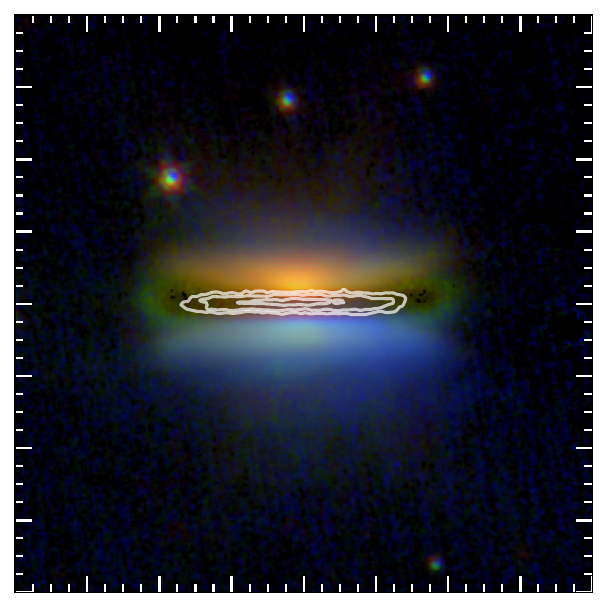}
    \caption{A red, green, and blue (RGB) composite image of the Flying Saucer. The RGB channels correspond to the JWST F460W-, F200W-, and F115W- band images, respectively. An asinh stretch is applied to enhance low-surface-brightness features, and each channel is normalized by the integrated disk flux. ALMA band 7 dust continuum contours at 5, 10, 20$\sigma$ are overlaid in white. White ticks are shown every 1\arcsec, with minor ticks every 0.25\arcsec.}
    \label{fig: rgb}
\end{figure}

\begin{figure*}
    \centering
    \includegraphics[width=0.98\linewidth]{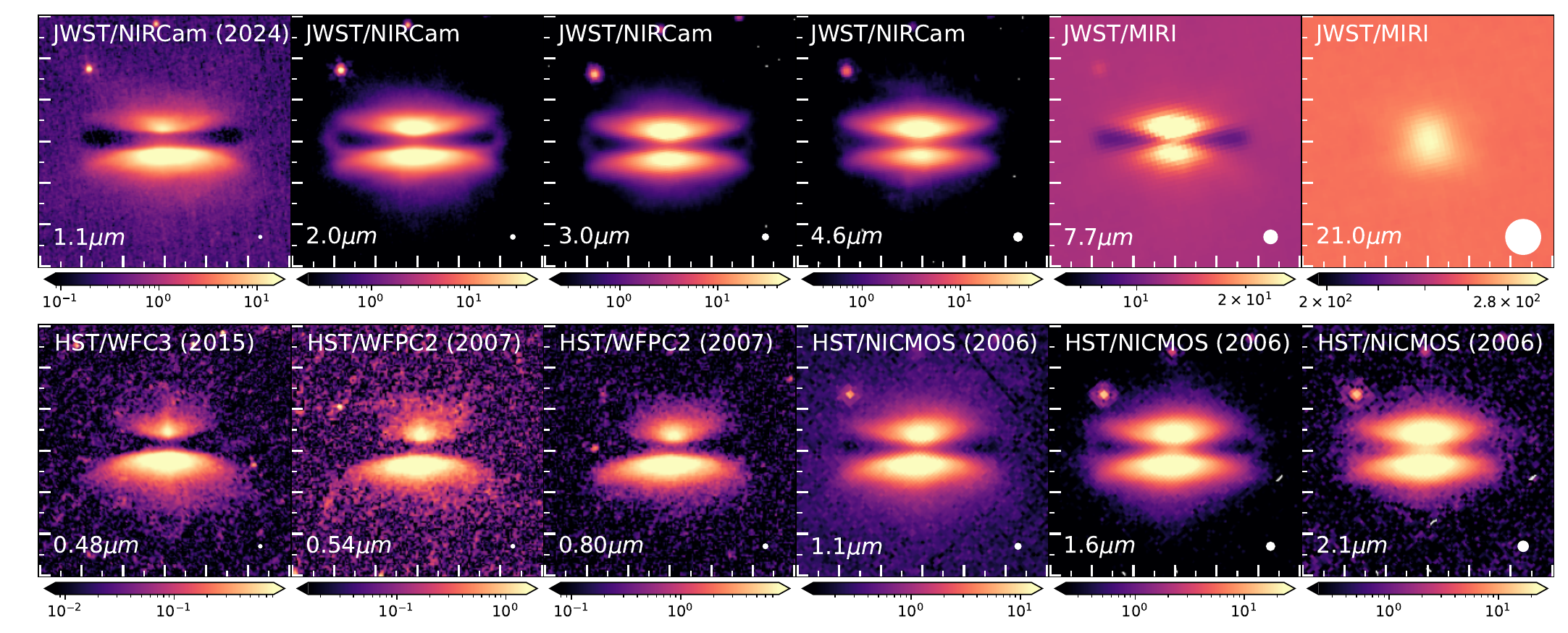}
    \caption{Gallery of broadband images of the Flying Saucer obtained with JWST(top row) and HST(bottom row), arranged in increasing wavelength order. All panels show a $6\arcsec \times 6\arcsec$ field of view. ick marks are shown at intervals of 1\arcsec, with minor ticks every 0.5\arcsec. The central wavelength and observing epoch are indicated at the top of each panel. A logarithmic color scale is used, and the units are in MJy/sr. The size of PSFs is plotted at the right corner. }
    \label{fig:gallery}
\end{figure*}

We observed the Flying Saucer as part of the JWST cycle 2 GO program 4290 (PI: F. Ménard, K.R. Stapelfeldt), with NIRCam and MIRI in consecutive visits on 2024 April 9. The NIRCam observations were made in four different filters F115W, F200W, F300W, and F460W, for a total integration time of 859 s in each. The MIRI observations were made in the F770W and F2100W filters, with integration times of 426 and 495 s, respectively. An archival F1280W image of the target was expected to come from GTO program 1282 but failed in execution and was not repeated.  
The *i2d* images from the standard JWST pipeline version 1.14 were retrieved from the STScI/MAST archive. 
A red, green, and blue composite image of the reduced JWST data is shown in Fig.~\ref{fig: rgb}. The reduced data for each wavelength are shown in Fig.~\ref{fig:gallery} in a logarithmic stretch. 

As a complement, we include NIRSpec observations from GO program 5299 (PI: J. Bergner). The data were obtained on 2024 August 25, using the G235H and G395H gratings, covering wavelength ranges of 1.66-3.17$\mu$m and 2.87-5.27$\mu$m, respectively, at a spectral resolution of $R=2700$. We retrieved the level 3 *s3d* products reduced with version 1.18.0 pipeline from the STScI/MAST archive. These observations were presented by \citet{Dartois2025}, where further details can be found.

\subsection{HST} \label{subsec:obs-hst}
The Flying Saucer was observed with three different Hubble
Space Telescope (HST) cameras, each at a different epoch (see Table 1).  GO Program 10603 (PI: D. Padgett) imaged it in 2006 with the NICMOS NIC2 camera in the F110W filter for 1024 s, and in the F160W and F205W filters for 768 s each.  GO 10603 also imaged it with the WPFC2 PC1 camera for 1200 s in the F555W filter and 800 s in the F814W filter in 2007.  A blue F475W image was made with the WFC3 UVIS instrument in Program 13643 (PI: G. Duche\^ne) in 2015 for 2190 s.  All the HST images were split in two to reject cosmic rays.  Since the three sets of HST images were all taken with different pixel scales and sky orientations versus the JWST NIRCam images, they were rotated and resampled to orient the disk major axis (position angle 93$^{\circ}$) along the image X-axis and with the pixel scale of the NIRCam blue cameras - as was already done for the JWST MIRI images.

It was also necessary to register the set of JWST and HST images so that the center of the disk was commonly aligned in all of them.  To do this, the photometric centroid of the upper nebula was measured inside the faintest isophotal aperture that still distinctly separated the upper nebula from the lower nebula.  In the rotation and resampling step above, the images were also translated such that the centroid of the upper nebula was aligned across all 12 broadband images. 
The reduced images are shown in Fig.~\ref{fig:gallery}. 

\subsection{ALMA} \label{subsec:obs-alma}
ALMA Band 7 observations of the Flying Saucer were obtained as part of project 2022.1.00742.S (PI: F. Ménard) on 2023 June 7, with an on-source integration time of 14 minutes. The data were calibrated with the ALMA pipeline in CASA V6.5.4.9 and imaged using Briggs weighting with a robust parameter of 1.0. The observation has been published in \citet{Martinien2026}, where further details on data reduction can be found.

We recentered the image on its intensity-weighted center and derotated it by $3.3^\circ$, as estimated from an ellipse fit to the contour. The millimeter emission is overlaid as contours in Fig.~\ref{fig: rgb} and in log-stretch in Fig.~\ref{fig:mcfost_ref}. The millimeter emission nicely aligns with the scattered-light, concentrating onto the midplane. 


\subsection{TESS \& K2} \label{subsec:obs-tess}
Like all young stars, the Flying Saucer is photometrically variable.  Because this variability is potentially connected to the wavelength-dependent brightness ratio between the two lobes of nebulosity, we examined the available space-based time series photometry of the system. The source was observed by the K2 mission Campaign 2 in the fall of 2014 for a 90 day stare at 30 minute cadence.  We retrieved the level 3 K2 lightcurve, which \cite{Cody2018} classified as quasi-periodic.  In TESS Cycle 7, we proposed for a follow-up time series to see if the source had a persistent mode of variability.  Mini-proposal 7060 (PI: K. Stapelfeldt) was accepted, and the target was observed at 2 minute cadence for 28 days in Sector 91, spring 2025. Unfortunately, the TESS spacecraft experienced a safe mode during these observations, which resulted in the loss of 4 days of the time series.  We retrieved level 3 presearch data conditioning simple aperture photometry (PDCSAP) TESS Light curves from MAST archive (Observation ID: tess-s0091-3-3).

\begin{figure}
    \centering
    \includegraphics[width=0.98\linewidth]{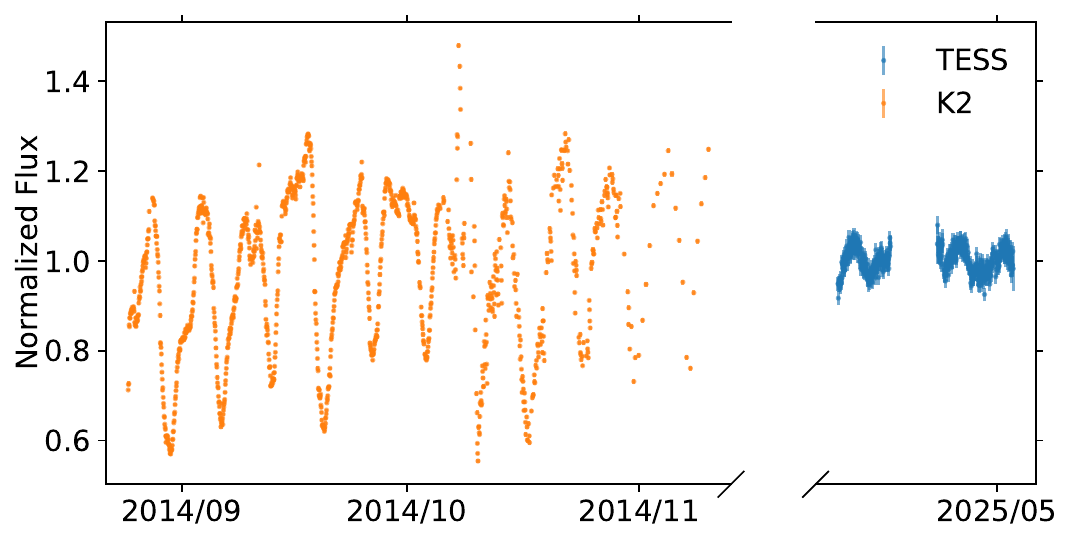}
    \caption{TESS and K2 Lightcurves. The lightcurves are normalized with their median value respectively. }
    \label{fig:tess-k2}
\end{figure}

For both lightcurves, we removed NaN values and rejected outliers using the default 5$\sigma$-clipping algorithm implemented in lightkurve, which is designed for Kepler and TESS data analysis \citep{Lightkurve}. Then we rebinned it by every hour, and normalized the curve to the median value, shown in Fig.~\ref{fig:tess-k2}.  The K2 lightcurve has a variation amplitude of about 30\%, while the TESS lightcurve variation amplitude was only $10\%$.  This may be due to intrinsic astrophysical changes in the source between the two epochs, or could also be due to the different wavelengths of the two observations ($0.65\mu m$ for K2, $0.8\mu m$ for TESS).



\section{Results} \label{sec:result}
\subsection{Overview} \label{subsec:overview}
From 0.5$\mu$m to 7.7$\mu$m, the Flying Saucer is well resolved and consistently appears as two reflection nebulae separated by a dark lane. This characteristic morphology is preserved across this wavelength range, despite changes in the surrounding emission and wavelength-dependent relative brightness variation between the two nebulae. We characterize the disk morphology following the definitions introduced by \citet{Duchene2024}. Table.~\ref{tab:disk-morph} briefly summarizes the definitions and presents the result for JWST and HST observations.

The disk exhibits silhouette features at multiple wavelengths. At 7.7$\mu$m, the deep absorption is clearly seen against the PAH emission from the $\rho$~Ophiuchi nebula. The 1.1$\mu$m silhouette feature is weaker (see Fig.~\ref{fig:obs-epoch} and the 0.48$\mu$m feature is marginal. The detailed analysis and modeling will be detailed in \ref{sect: silhouette} and \ref{sect: silhouette_model}. 

At 21$\mu$m, the disk no longer exhibits a clear two-nebulae morphology and the dark lane is not visible. Instead, it appears as a single, compact emission feature due to coarser angular resolution at this wavelength, against a strong surrounding sky emission. Following the method described by \citet{Tazaki2025}, we fit the point-spread function (PSF) to the vertical cut along the minor axis and find that the source remains resolved, allowing a tentative separation of the two nebular components.

For the first time, we detect a bright, curved lining along both outer radial edges at 2 and 3$\mu$m, connecting the two reflection nebulae across the midplane (See Fig.~\ref{fig: rgb},~\ref{fig:gallery}). 
The modeling is present in Section~\ref{sect:lining} and tentative physical origins are discussed in Section~\ref{sect: lining_discussion}. 

The relative brightness of the two nebulae varies with wavelength. The bottom nebula is brighter at wavelengths shorter than 2$\mu$m, while the top nebula becomes brighter at longer wavelengths.
Fig.~\ref{fig:obs-epoch} compares multi-epoch images obtained with HST and JWST and reveals temporal variations in nebular brightness at 1-2$\mu$m. Despite this variability, the bottom nebula consistently remains the brighter one at wavelengths shorter than 2$\mu$m across all epochs.
Lateral asymmetry is also present in the disk and varies between epochs.
In addition, three field sources are detected above the disk and clearly moved relative to the disk. These sources serve as useful astrometric reference points for proper motion analysis. 


\begin{table*}[]
    \centering
    \caption{Disk Morphological Properties of the Flying Saucer}
    \label{tab:disk-morph}
    \small
    \setlength{\tabcolsep}{3pt}
    \begin{tabular}{l c c c c c c c c c c}
    \hline
    \hline
    Instrument    & Filter  & $\lambda$  &  Date   & $d_{\rm neb}$ &  $R_{\rm peak}$  & Flux ratio &  FWHM$_{\rm top}$ & FWHM$_{\rm bot}$ &  FW10\%$_{\rm top}$ &  FW10\%$_{\rm bot}$\\
                  &         & ($\mu$m)  &         &  ($\arcsec$)  &              &   & ($\arcsec$)      & ($\arcsec$)      & ($\arcsec$) & ($\arcsec$)\\
    \hline
    HST/WFC3      & F475W   & 0.48  & 2015/08/07 & 0.66$\pm$0.01 & 0.39 & 0.38 & 0.31 & 0.84 & 1.64 & 2.29  \\
    HST/WFPC2     & F555W   & 0.54  & 2007/03/27 & 0.69$\pm$0.01 & 0.43 & 0.47 & 0.50 & 0.96 & 1.36 & 2.42  \\
    HST/WFPC2     & F814W   & 0.80  & 2007/03/27 & 0.69$\pm$0.01 & 0.50 & 0.44 & 0.53 & 0.99 & 1.71 & 2.48 \\
    HST/NICMOS    & F110W   & 1.1   & 2006/05/17 & 0.68$\pm$0.01 & 0.55 & 0.60 & 0.68 & 0.87 & 2.02 & 2.39 \\
    HST/NICMOS    & F160W   & 1.6   & 2006/05/17 & 0.68$\pm$0.01 & 0.70 & 0.73 & 0.78 & 0.93 & 2.20 & 2.45 \\
    HST/NICMOS    & F205W   & 2.1   & 2006/05/17 & 0.66$\pm$0.01 & 0.96 & 0.94 & 0.84 & 0.96 & 2.23 & 2.51 \\
    JWST/NIRCam   & F115W   & 1.1   & 2024/04/09 & 0.65$\pm$0.02 & 0.31 & 0.37 & 0.65 & 1.08 & 2.20 & 2.54 \\
    JWST/NIRCam   & F200W   & 2.0   & 2024/04/09 & 0.63$\pm$0.01 & 0.97 & 0.67 & 0.59 & 0.93 & 1.80 & 2.42 \\
    JWST/NIRCam   & F300M   & 3.0   & 2024/04/09 & 0.61$\pm$0.02 & 1.77 & 1.12 & 0.56 & 0.90 & 1.74 & 2.39 \\
    JWST/NIRCam   & F460M   & 4.6   & 2024/04/09 & 0.59$\pm$0.01 & 2.58 & 1.78 & 0.65 & 0.81 & 1.83 & 2.26 \\
    JWST/MIRI     & F770W   & 7.7   & 2024/04/09 & 0.55$\pm$0.02 & 2.00 & 1.28 & 0.84 & 0.90 & 2.02 & 2.36 \\
    JWST/MIRI    & F2100W  & 21.0  & 2024/04/09 & 0.58$\pm$0.01  & 1.0 & 0.93 & - & - & - & -\\
    \hline
    \end{tabular}
    \begin{minipage}{\linewidth}
    \tablecomments{The morphological characteristics primarily follow the definitions in \citet{Duchene2024} and \citet{Tazaki2025}. For each image, vertical intensity cuts are extracted across the disk and fitted with polynomials of an order of 5 to 10 to trace the spines of the top and bottom nebulae. The separation between the two nebulae, or dark lane thickness $d_{\rm neb}$, is defined as the minimum distance between the two spines. The final value of $d_{\rm neb}$ is taken as the mean obtained from the different polynomial orders, while the corresponding standard deviation is adopted as its uncertainty. FWHM${\rm top}$ and FWHM${\rm bot}$ denote the full width at half maximum of the intensity profiles measured along the top and bottom spines, respectively. $\mathrm{FW10\%}{\rm top}$ and $\mathrm{FW10\%}{\rm bot}$ are defined as the widths enclosing the top 10\% of the corresponding intensity profiles. For the F2100W image, the $d_{\rm neb}$ is characterized by fitting two point spread function cuts along the minor axis.
    The peak ratio $R_{\rm peak}$ is defined as the ratio of the brightest pixel intensities in the two nebulae, while the flux ratio corresponds to the ratio of the nebula-integrated fluxes.
    }
    \end{minipage}
    
\end{table*}

\subsection{Photometry}
\label{sect: photometry}
For all images, we subtract the surrounding emission estimated as the mean flux within an annulus of [3.7,4]\arcsec. We then integrate the disk flux within a circular aperture of radius 3\arcsec. The resulting photometry is summarized in Table.~\ref{tab:photo}. 

Previous near-infrared photometry reported no dramatic brightness variations on timescales of $\sim$3 years \citep{Grosso2003, Cutri2003}. The Two Micron All Sky Survey (2MASS) magnitudes reported by \citet{Grosso2003} corresponds to fluxes of $0.68\pm0.04$~mJy ($J$), $1.25\pm0.05$~mJy ($H$), $1.70\pm0.04$~mJy ($Ks$), which are consistent with our measurements within the $\sim$10\% variability amplitude inferred from the TESS light curve. The fluxes reported by \citet{Cutri2003} are systematically higher ($0.90\pm0.06$, $1.66\pm0.07$, $1.97\pm0.09$~mJy, for $J$,$H$,and $Ks$, respectively). In the mid-infrared, the fluxes reported with Spitzer by \citet{Pontoppidan2007} are also higher than those derived from our JWST data. This discrepancy is likely due to contamination from surrounding diffuse emission contribution, which is more difficult to separate from the disk emission in lower angular resolution observations. 

To further quantify the variability amplitude suggested by these photometric differences, we analyze the K2 light curve, which covers a longer time span. Using the variability metrics from \citet{Cody2014}, we derive a periodicity parameter $Q = 0.31$ and asymmetry parameter $M=0.61$, classifying the source as a quasi-periodic dipper. A Lomb–Scargle periodogram yields a dominant period of $P=4.89$~days. Such variability is commonly attributed to recurring passage of occulting material in the inner disk, or to illumination inhomogeneities such as beaming from stellar accretion hot spots \citep{Wood1998}.

\begin{table}[]
    \centering
    \caption{Photometry of Flying Saucer}
    \label{tab:photo}
    \begin{tabular}{c c c}
    \hline
    \hline
    Date & $\lambda$ &  Flux \\
         & ($\mu$m) &  (mJy) \\ 
    \hline 
    2015/08/07     & 0.48 & 0.034 $\pm$ 0.001 \\
    2007/03/27     & 0.54 & 0.075 $\pm$ 0.002 \\
                   & 0.8  & 0.377 $\pm$ 0.008\\
    2006/05/17     & 1.1 &  0.67 $\pm$ 0.02\\
                   & 1.6 &  1.51 $\pm$ 0.05\\
                   & 2.1 &  1.87 $\pm$ 0.06\\
    2024/04/09     & 1.1 & 0.62 $\pm$ 0.01 \\
                   & 2.0 & 1.86 $\pm$ 0.04 \\
                   & 3.0 & 1.63 $\pm$ 0.03 \\
                   & 4.6 & 1.63 $\pm$ 0.07 \\
                   & 7.7 & 0.69 $\pm$ 0.01 \\
                   & 21.0 & 1.34 $\pm$ 0.01 \\
    \hline
    \end{tabular}
    \begin{minipage}{\linewidth}
    \tablecomments{The flux is integrated within a round aperture with radius 3\arcsec after the background removal. The background is estimated within an annulus of [3.7, 4]\arcsec. HST flux uncertainties are $1.8\%$ for WFC3, $2\%$ for WFPC2 broad-band, and $3\%$ for NICMOS according to the handbook.}
    \end{minipage}
\end{table}

\subsection{Proper Motion}
\label{sect: proper_motion}
\begin{figure}
    \centering
    \includegraphics[width=0.95\linewidth]{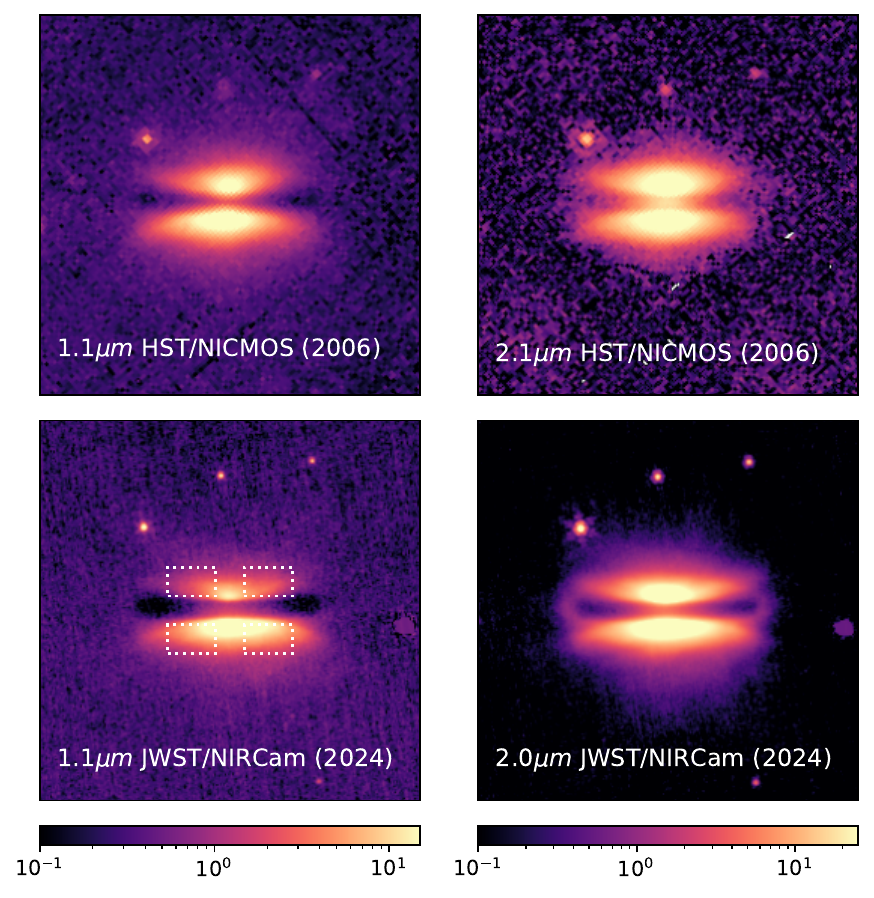}
    \caption{HST(2006) and JWST(2024) at 1.1 and 2.0$\mu$m, using a log-scale color stretch to highlight the faint substructures. The same wavelength shares the same colorbar in MJy sr$^{-1}$. The dashed rectangles in the JWST 1.1$\mu$m panel indicate the regions used to measure the lateral asymmetry. All panels are displayed in $8\arcsec\times 8\arcsec$.}
    \label{fig:obs-epoch}
\end{figure}

The JWST and HST observations are separated by 18 years, and both include 1.1$\mu$m and 2.0$\mu$m broadband imaging (Fig.~\ref{fig:obs-epoch}). Three field sources are detected in both epochs and exhibit clear relative motion with respect to the Flying Saucer. 

We fitted 2D Gaussian to determine the locations of the sources, and calculated the relative shift between two epochs at both wavelengths. We find a consistent offset of $374\pm38, 408\pm 44$ and $408 \pm 40$~mas for three sources, all in the same directions and consistent within the uncertainties. Considering these sources as stationary objects, the measured relative shifts correspond to a proper motion of $-6.9\pm0.5$ mas~yr$^{-1}$ in R.A. direction, and $-20.7\pm 0.8$ mas~yr$^{-1}$ in the decl. direction, for the Flying Saucer.

From the Gaia archive, we retrieved 51 sources that are located within 20$\arcmin$ around the Flying Saucer and $140\pm10$~pc away, as representatives of the $\rho$~Ophiuchi cloud. These sources yield a mean proper motion of $-7.1\pm1.7$ mas~yr$^{-1}$ in R.A. and $-26.2\pm 1.6$ mas~yr$^{-1}$ in the decl. direction, consistent with 
Gaia EDR3 for young objects in the region \citep{Grasser2021}. 
Therefore, we conclude that the Flying Saucer is comoving with the nearby sources in $\rho$~Ophiuchi. 

\subsection{Flux Ratio}
\label{sect: flux_ratio}
\begin{figure}
    \centering
    \includegraphics[width=0.95\linewidth]{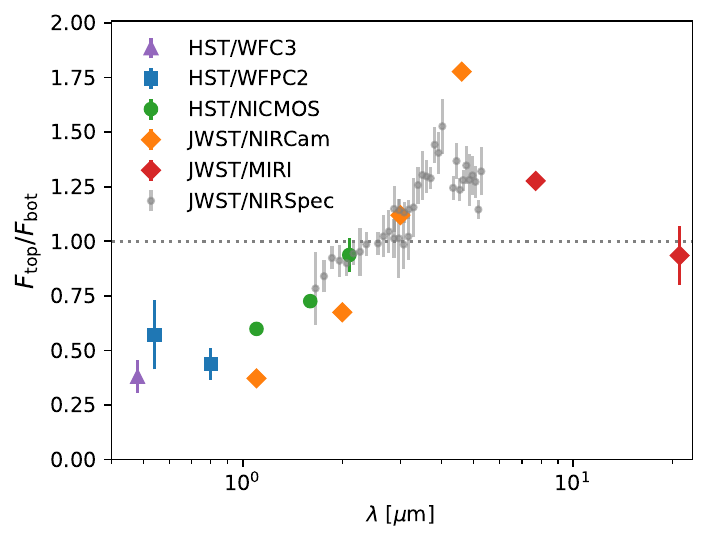}
    \caption{A flux ratio between the top and bottom nebulae as a function of wavelength. Symbols denote measurements from different instruments, as indicated in the legend. The horizontal dotted line marks a flux ratio of unity.}
    \label{fig:flux_ratio}
\end{figure}

We integrate the flux of the top and bottom nebulae separately within rectangular apertures of a height of $1.2\arcsec$ and a width of $4.3\arcsec$. We then compute the flux ratio between the top and bottom nebulae. The results are shown in Fig.~\ref{fig:flux_ratio}, where measurements from different epochs are overplotted using distinct colors and markers. The flux ratio is approximately 0.5 at wavelengths shorter than $\sim$1$\mu$m, increases monotonically with wavelength, crosses 1 at $\sim$2$\mu$m, and reaches a maximum value of $\sim$1.8 at 4.6$\mu$m. At longer wavelengths, the ratio decreases again and approaches 1 at 21$\mu$m, where the two nebulae become equally bright again. While the disk-integrated flux remains relatively stable across epochs, the nebula-integrated flux exhibits larger variability comparing HST and JWST images. 

We also include JWST/NIRSpec data obtained approximately 4 months after broadband imaging. We discarded the spectral channels in which the disk is not fully covered, rebinned the remaining data by 0.1$\mu m$, recentered and derotated the images. We applied the same measuring process as for broadband images, and overplotted the results in Fig.~\ref{fig:flux_ratio} as the gray points. Although the uncertainties are larger towards the spectral edges, the NIRSpec measurements follow the same trend as the the broadband imaging and confirm that the flux ratio reversal occurs at 2-3$\mu$m. With previously reported peak brightness reversal between $H$ and $K$ bands by \citet{Grosso2003}, the switching wavelength is stable across epochs. 


\begin{figure}
    \centering
    \includegraphics[width=0.95\linewidth]{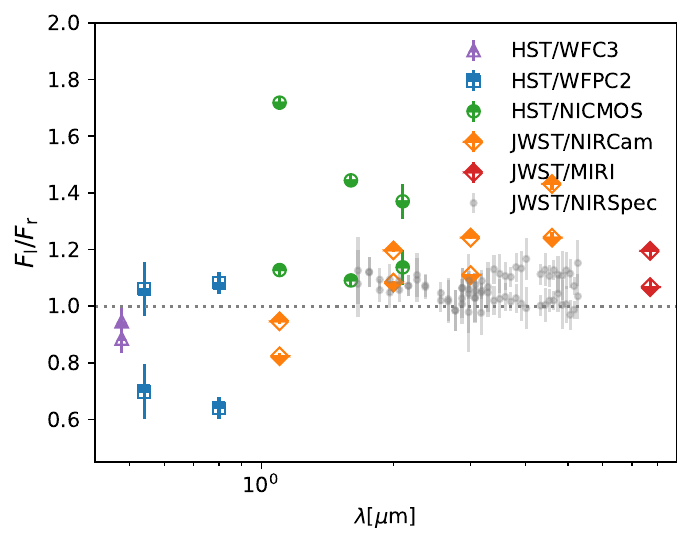}
    \caption{Lateral asymmetry measured from flux ratios between the left and right sides of reflection nebulae as a function of wavelength. The asymmetry is quantified using four rectangular apertures of $1.0\arcsec\times0.6\arcsec$, offset by $0.3\arcsec$ from the image center. Half-filled markers distinguish flux ratios measured in the top and bottom nebulae.}
    \label{fig: asymmetry}
\end{figure}

\textit{Lateral asymmetry.} To quantify the lateral asymmetry in the disk, we define 4 rectangular apertures of a height of $1.0\arcsec$ and a width of $0.6\arcsec$, offset by $0.3\arcsec$ to the left, right, above, and below the image center, shown in Fig.~\ref{fig:obs-epoch}. These regions are chosen to cover the larger separations and avoid the dark lane. The lateral asymmetry is then measured as the flux ratio between the left and right sides within the top and bottom reflection nebulae, denoted as $F_{\rm ul}/F_{\rm ur}$ and $F_{\rm ll}/F_{\rm lr}$ respectively. The results are presented in Fig.~\ref{fig: asymmetry}, where half-filled markers indicate measurements from the top and bottom nebula. Measurements from the NIRSpec data are included, but separate symbols are not used because the two asymmetry estimates are consistent within uncertainties, and indicate an overall left-right symmetry.
We tested the robustness of the measurements by varying both the aperture size and their radial offsets. The derived flux ratios remain unchanged within the uncertainties. Overall, the lateral $F_{\rm l}/F_{\rm r}$ values lies between 0.5 and 2, comparable to values reported for HH~30\citep{Watson2007}. The most pronounced asymmetries occur in the top nebula at $0.5$ and 0.8$\mu$m, where the right side is brighter; and in the bottom nebula at 1.1$\mu$m, where the left side is brighter. 

Notable variability is present in lateral asymmetry. In the JWST observations, the $F_{\rm ul}/F_{\rm ur}$ is systematically higher than $F_{\rm ll}/F_{\rm lr}$ at all wavelengths, whereas HST data show the opposite between 1 and 2$\mu$m. Moreover, the lateral asymmetry $F_{\rm l}/F_{\rm r}$ correlates with the top–bottom flux ratio $F_{\rm top}/F_{\rm bot}$. When the top nebula relatively brightens with wavelength, the left side also becomes relatively brighter. An opposite behaviour is seen in the HST data, as the right side becomes brighter. 

We do not adopt the lateral asymmetry parameter $\delta_{\rm spine}$ introduced by \citet{Villenave2024}, as its measurement depends on precise image centering and midplane orientation. This quantity is straightforward to evaluate in simulations, but in observations, these parameters can be less well constrained, especially in the presence of intrinsic disk asymmetries.

\subsection{Major Axis Profiles: Silhouette and Lining}
\label{sect: silhouette}

\begin{figure}
    \centering    
    \includegraphics[width=0.95\linewidth]{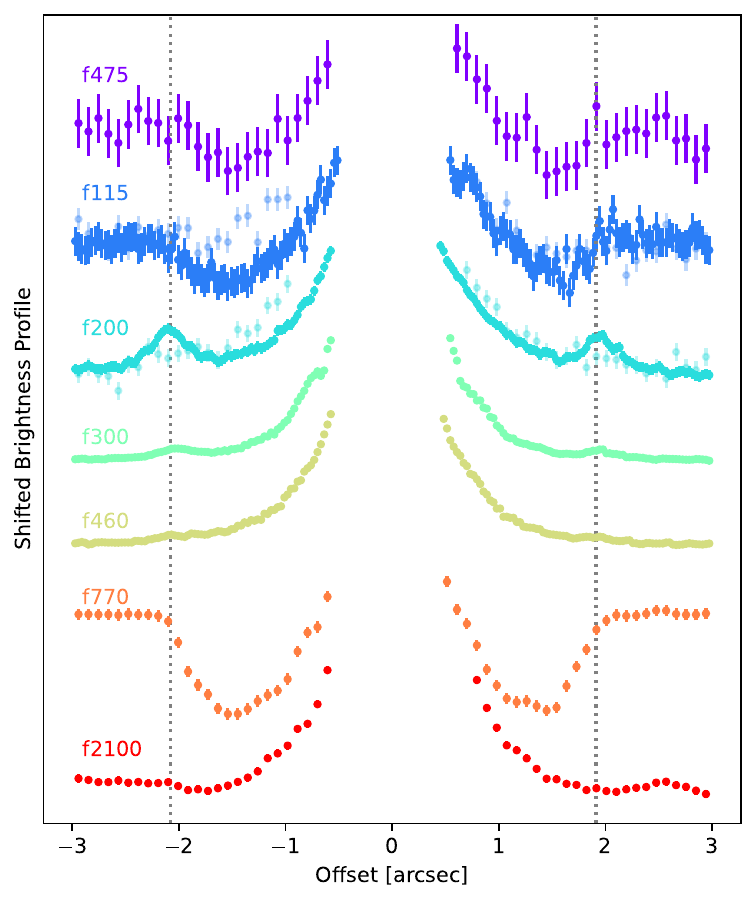}
    \caption{Brightness profiles through the major axis for images observed with JWST. The profiles are normalized to its brightness at 0.7\arcsec and shifted for clarification. The dashed lines indicate the lining peaks. HST profiles are overplotted at 1.1 and 2.0$\mu$m as lighter symbols. The central region is excluded as it is dominated by PSF convolution.}
    \label{fig:majorcut}
\end{figure}

We analyze the wavelength-dependent disk-edge structure using intensity profiles extracted along the disk midplane from the JWST broadband images. Compared to spectroscopic imaging, the broadband data provide higher angular resolution for quantitative characterization. Selected profiles are shown in Fig.~\ref{fig:majorcut}, including F475W data obtained with HST. The uncertainties are estimated from the standard deviation of the flux measured in a surrounding annulus. All the profiles are normalized to their brightness at $0.7\arcsec$ and shifted for better visual comparison. The center peak is omitted because it is dominated by PSF convolution effects and is not relevant to the edge feature discussed here. 


The disk exhibits clear absorptive silhouette features at 1.1$\mu$m and 7.7$\mu$m with different depths. At 0.47$\mu$m, the silhouette is only marginally detected, while at 21$\mu$m a shallow tentative absorption feature is present. \citet{Dartois2025} has recently reported this disk to be seen in silhouette over H2 background emission at $17.2\mu m$ with MIRI medium resolution spectroscopy. 
To quantify the absorption features for F115W and F770W data, we first shift each intensity profile so that its minimum is zero. This step is physically motivated by the expectation that the disk midplane is optically thick ($\tau\gg1$), so that ideally, the observed intensity at the deepest point of the silhouette would be dominated by foreground emission. So bringing the flux to zero removes the foreground emission and leaves the profile with background emission. Then we normalize the profile by the surrounding emission, estimated separately on both sides of the disk within [3.7\arcsec, 4\arcsec], where the signal is dominated by the background. The normalized profiles are shown in Fig.~\ref{fig:sil_edge}. 
We note, however, that this procedure is approximate. Although the scattered light from the disk surface does not directly affect the brightness profile through the midplane, PSF convolution redistributes the surface flux into the midplane and prevents the observed profiles from reaching zero flux. This effect is particularly important at longer wavelength 7.7$\mu$m , where the PSF is broader. Later in Section.~\ref{sect: silhouette_model} we apply the same normalization procedure to both the observations and the convolved model images fitted by forward modeling, ensuring that the resulting edge characteristics remain quantitatively meaningful. 

For the absorptive feature, we fit the silhouette region with an eighth-order polynomial to the normalized profiles, and define two characteristic quantities: (a) the edge position $r_{50\%}$, defined as the radius at which 0.50 is reached; (b) the normalized silhouette edge width $\Delta r/r$, quantifies by the normalized radial separation between the 25\% and 75\% background flux levels, $\Delta r/r = 2\cdot(r_{75\%} - r_{25\%})/(r_{75\%} + r_{25\%})$. Smaller values of $\Delta r/r$ correspond to sharper edges. $\Delta r/r$ is not readily fitted on the right side at 1.1$\mu$m and hence not given. 
The typical fitting uncertainty is $0.01\arcsec$ on the $r_{50\%}$, and 0.01 on the $\Delta r/r$, by varying the order of polynomial fitting. The same measurement procedure will be applied to PSF-convolved model images. Although the pixel-by-pixel uncertainty is larger at $1.1\mu m$, the fitting of $r_{50\%}$ is quite stable. Both the measured $r_{50\%}$ and $\Delta r/r$ remain approximately constant from $1.1$ to $7.7\mu m$ within the uncertainties. The left radius is slightly larger than the right one, consistent with the fact that the disk shows a slightly brighter left side (see Fig.~\ref{fig: asymmetry}).

In contrast, 2.0 and 3.0$\mu$m highlight a lining feature, indicative of enhanced scattering rather than absorption. 
To characterize the lining feature, we decompose the profiles at F200W and F300W as a second-order polynomial plus a Gaussian peak between the region [1.4, 2.8]\arcsec. The Gaussian peak position is considered as the lining location, and its strength is quantified by the ratio between the Gaussian peak and the underlying continuum. The fitted results are summarized in Table.~\ref{tab:silhouette}. 

For comparison, we overplot HST brightness profiles at 1.1 and 2.0$\mu$m as lighter symbols in Fig.~\ref{fig:majorcut}. Neither the silhouette feature at 1.1$\mu$m nor the lining at 2.0$\mu$m was detected with HST, due to lower spatial resolution and higher noise compared to JWST.

In addition, we quantified $r_{95\%}$, defined as the radius at which 0.95 is reached, as an estimate of the outermost detectable silhouette edge and to compare with the lining position. At $1.1\mu m$, we obtain $r_{95\%} = 2.1\arcsec$ on the left and not readily measured on the right side. At $7.7\mu m$, $r_{95\%}$ is robustly measured at 2.15$\arcsec$ and 2.07$\arcsec$ for the left and right sides. 
The lining peak coincides with $r_{95\%}$, indicating that both features trace the same disk boundary. The lining feature is more pronounced at 2$\mu$m. Its strength is comparable on both sides.

\begin{table}[]
    \centering
    \caption{Characteristic quantities of the silhouette edge (top) and the lining feature (bottom).}
    \label{tab:silhouette}
    \begin{tabular}{c c c c c}
    \hline
    \hline
    Filter   &  \multicolumn{2}{c}{Left} & \multicolumn{2}{c}{Right} \\
    \hline
    \multicolumn{5}{c}{Silhouette}\\
    &$r_{50\%}$ & $\Delta r/r$ &$r_{50\%}$ & $\Delta r/r$  \\ 
    F115W   & 1.90 & 0.11 & 1.70 & - \\
    F770W   & 1.90 & 0.13 & 1.77 & 0.16 \\
    \hline  
    \multicolumn{5}{c}{Lining}\\
    & Position & Strength  & Position  & Strength \\
    F200W    & 2.10 & 2.74 & 1.96 & 2.03 \\
    F300W    & 2.05 & 1.14 & 1.95 & 1.26 \\
    \hline
    \end{tabular}
    \begin{minipage}{\linewidth}
    \tablecomments{
        For the silhouette edge characterization, the edge position $r_{\rm 50\%}$ is defined as the radius where the normalized intensity reaches 0.50. The edge width $\Delta r/r = 2\cdot(r_{75\%} - r_{25\%})/(r_{75\%} + r_{25\%})$ represents the normalized radial separation between the 25\% and 75\% levels, with smaller values indicating sharper edges. The typical fitting error is 0.01\arcsec for $r_{50\%}$ and 0.01 for $\Delta r/r$ by varying the order of the polynomial fitting. For the lining feature, the position is defined as the center of the Gaussian component fitted to the profile, and its strength is given by the ratio of the Gaussian amplitude to the underlying continuum level. }
    \end{minipage}
\end{table}

\section{Radiative Transfer Modeling} \label{sec:model}
\subsection{Motivation}
The radiative transfer model in this work has three primary goals. To address them, we construct physically plausible models that provide a reasonable match to the three main observed features. The model is not intended to provide a unique or global best fit to the full multiwavelength dataset. An exhaustive exploration of all parameters using Markov Chain
Monte Carlo is beyond the scope of this study.

First, we investigate the wavelength-dependence of the dark-lane thickness, a classic diagnostic of the vertical dust distribution in edge-on disks that provides direct constraints on dust grain growth and vertical mixing. By comparing the observed dark-lane thickness across optical to mid-infrared wavelengths with radiative transfer models, we aim to quantify the degree of vertical dust mixing in the Flying Saucer disk.

Second, we aim to investigate the physical origin of the flux ratio reversal between the two nebulae. Flux ratio reversals have also been identified in other edge-on disks, such as IRAS~04302, at longer wavelengths between 12.8 and 21$\mu$m and a misaligned inner disk has been proposed as a plausible explanation \citep{Villenave2024}. Here, we test whether such a misalignment can also reproduce a flux ratio reversal at shorter, near-infrared wavelengths, identify the key parameters controlling the emergence of the reversal. 

Third, we aim to interpret the absorptive silhouette and scattering lining features. Using the multi-wavelength radiative transfer model, we investigate whether these two features can be explained by a common disk structure and assess what constraints the wavelength-dependent absorption or scattering places on the dust density distribution and grain sizes in the disk.

\subsection{Model Description}
\label{sect: ref_model}
We used the radiative transfer code MCFOST \citep{Pinte2006, Pinte2009} to model the Flying Saucer. Here we introduce a "reference model", intended as a physically plausible configuration that reproduces the main observed characteristics and serves as the baseline for later exploration on the dependence of the observables on disk geometry and dust properties. This model is not intended to represent a unique or formally best-fit solution.

To allow for a misalignment between the inner and outer disks, we adopted a two-zone disk structure. The inner disk extends from 0.1 to 3~au with a sharp outer edge. The outer disk extends from 5 to 300~au, and the edge profile follows $\Sigma(r)\propto r^{-1}\exp(-(r/r_c)^{2-\gamma})$, with a tapering index $\gamma=-0.5$ at outer edge. The radial gap between 3 and 5 au is introduced to avoid direct overlap between the two zones, and is also physically plausible for a disk composed of regions with different inclinations. This narrow gap has negligible effect on the scattered-light morphology (See Fig.~\ref{fig:mcfost_full} in the Appendix~\ref{app:model}). In the vertical direction, both zones adopt a Gaussian density structure, with the scale height given by $h(r) = h_{0} (r/r_0)^{\beta}$, where the scale height $h_0 = 12$au at $r_0=100$au and the flaring index $\beta=1.25$. 
The outer disk is viewed at an inclination of $i=87^\circ$ \citep{Dutrey2025}, with the bottom nebula tilted toward the observer. The inner disk is misaligned by $8^\circ$ relative to the outer disk at a viewing azimuth $\varphi = 90^\circ$, corresponding to an inclination of $i=85^\circ$ with the top nebula closer to the observer. These viewing parameters are adjusted to approximately reproduce the observed brightness ratios but are not uniquely constrained. $\varphi=90^\circ$ places the misalignment along the line of sight and defines a symmetric baseline geometry, such that no intrinsic left-right asymmetry is produced.
The dust masses of the inner and outer disks are set to $1.1\times10^{-6}$ and $3.2\times10^{-4} M_\odot$, respectively.

We assume a single dust composition throughout the disk, composed of 70\% of astrosilicate and 30\% graphite \citep{Draine1984} , to be consistent with the series of papers, although icy grains have been reported in the disk \citep{Dartois2025, Martinien2025}. The grain size distribution spans from 0.3 to 3000$\mu$m following $n(a)da\propto a^{p}da$ with $p=-3.5$. For grain size $a>a_{\rm mix}$, we introduced a parametric dust settling, with the scale height depending on the grain size $h(a) \propto (a/a_{\rm mix})^{\eta_{\rm settl}}$. Here we adopt $\eta_{\rm settl}=-0.5$ and $a_{\rm mix}=3\mu$m. The absorption and scattering properties are computed using the distribution of hollow spheres method \citep{Min2016}. We follow \citet{Grosso2003} and \citet{Pontoppidan2007} and set the central star $T_{\rm eff}= 3500$K and radius $R=1.03R_{\odot}$.
We adopted a distance of $d=140$~pc, corresponding to the mean distance of the $\rho$~Ophiuchi cloud \citep{Grasser2021}.  
We perform full scattering in 3D with $1.28 \times 10^7$ photons to compute the dust temperature and generate synthetic images $I_{\rm rt}(x,y)$ at 1.1, 2.0, 3.0, 4.6, 7.7, 21, and 870$\mu$m. 

\begin{figure*}
    \centering
    \includegraphics[width=\linewidth]{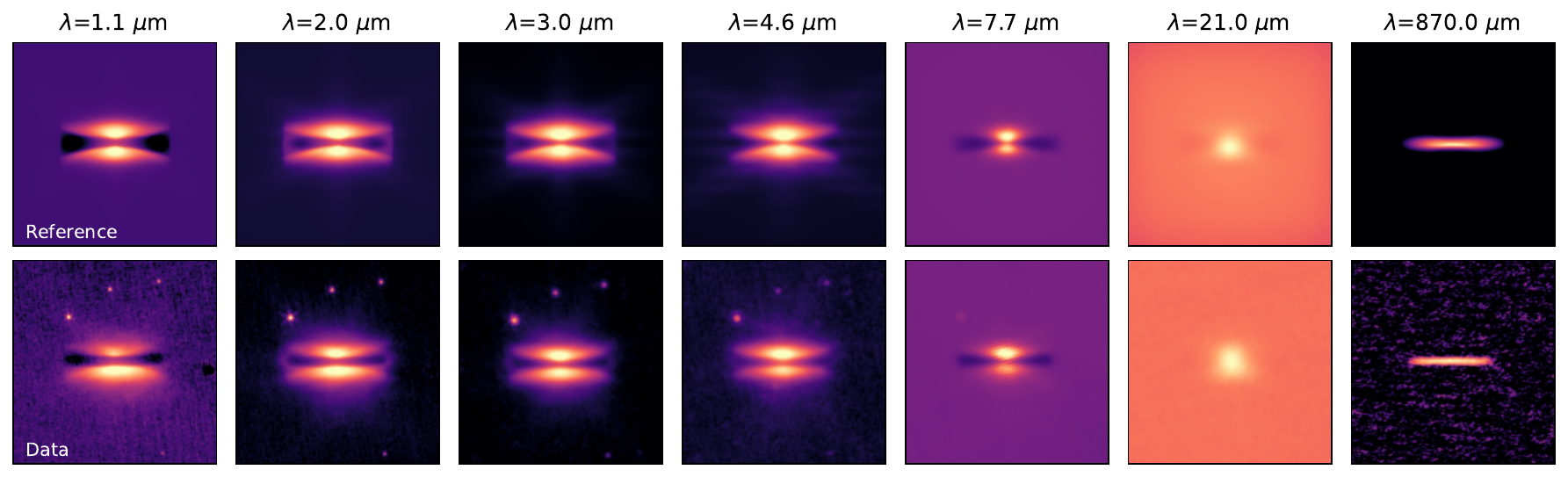}
    \caption{The simulated images for the reference model shown in logarithmic scale and the corresponding JWST and ALMA observations for comparison. The parameters of the reference model are fully described in Section~\ref{sect: ref_model}, and the fitted values for the surrounding environments and lining features are given in Table.~\ref{tab:baseline}. The simulated images are convolved with their corresponding PSFs. All panels display a field of view $8\arcsec \times 8\arcsec$.} 
    \label{fig:mcfost_ref}
\end{figure*}

In our implementation, MCFOST includes radiation from the central star and does not implicitly model extended environmental emission. To account for this indispensable contribution, we add three additional components to $I_{\rm rt}(x,y)$. 
First, we include background attenuation by assuming a spatially uniform extended emission level $I_{\rm bg}$ that is absorbed by the disk as $I_{\rm bg} e^{-\tau(x,y)}$, where $\tau(x,y)$ is the line-of-sight optical depth map computed by MCFOST. Second, we include a foreground term $I_{\rm fg}$, representing diffuse emission or extinction that is independent of the disk geometry or position on the sky.
Third, we account for forward scattering of extended nebular emission by dust grains near the disk surface. In the strongly forward-scattering regime, single scattering produces an intensity proportional to $\tau(x,y)e^{-\tau(x,y)}$, which peaks at $\tau\sim1$ and naturally produces a disk edge brightening. We model this effect using 
$I_{\rm fwd} = f\,I_{\rm bg}\,\tau(x,y)\,e^{-\tau(x,y)}$, where $f$ is a dimensionless scaling factor. This factor mainly encapsulates the dependence on dust albedo and the effective forward scattering lobe. The full derivation can be found in the Appendix~\ref{app: lining}.
This term is only included for F200W and F300W images, where the lining is observed. After adding these components, the images are convolved with their corresponding PSFs. The used PSF are from WebbPSF \citep{Perrin2014} for JWST wavelengths, and are generated using TinyTim \citep{Krist2011} for HST wavelengths. 
The final simulated image at the given wavelength is therefore: 
\begin{equation}
\begin{aligned}
I(x,y) = \Big[
    &I_{\rm rt}(x,y) + I_{\rm bg}\,e^{-\tau(x,y)} \\
    &+ I_{\rm fg} + f\,I_{\rm bg}\,\tau(x,y)\,e^{-\tau(x,y)}
\Big] \circledast {\rm PSF} .
\end{aligned}
\label{eq: intensity}
\end{equation}
At each wavelength, we then fit the parameters $(I_{\rm bg}, I_{\rm fg}, f)$ in Equation~\ref{eq: intensity} to the observations normalized to their peak value, and the results are summarized in Table.~\ref{tab:baseline}. The convolved synthetic images are shown in Fig.~\ref{fig:mcfost_ref}, which qualitatively reproduces the wavelength-dependent dark-lane thickness, the flux ratio reversal, and the lining features.
We defer the detailed fitting process and discussion to Section.~\ref{sect:lining}. 

\subsection{Results} 
\subsubsection{Dark Lane Thickness}
\begin{figure}
    \centering
    \includegraphics[width=\linewidth]{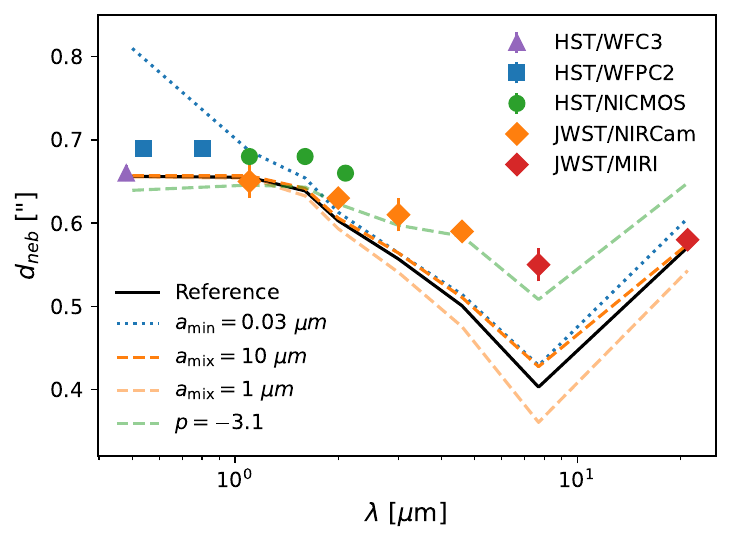}
    \caption{The dark-lane thickness $d_{\rm neb}$ as a function of wavelengths. The symbols show the measured values with error bars corresponding to the standard deviation obtained by varying the polynomial order used to trace the nebular spines. The overplotted curves show model predictions for different grain size distributions assuming the same dust composition and disk geometry. The reference model assumes dust species $a_{\rm min}=0.3\mu m$ and $a_{\rm mix}=3\mu m$ with $p=-3.5$. The colored lines are labeled by the parameter varied with respect to the reference model.}
    \label{fig:d_neb}
\end{figure}

The wavelength dependence of dark-lane thickness is an important indicator of the dust growth and vertical settling \citep{Watson2004, Duchene2024}. Fig.~\ref{fig:d_neb} summarizes the measured dark-lane thickness at observed wavelengths, overplotted with model predictions for different minimum grain size $a_{\rm min}$ and vertical mixing size $a_{\rm mix}$. When varying the minimum grain size $a_{\rm min}$, the total dust mass is adjusted accordingly, such that the modeled dark lane thickness at 1–2$\mu$m remains close to the observed value of $d_{\rm neb}\approx0.65\arcsec$.

Observationally, $d_{\rm neb}$ decreases from 0.7\arcsec at $\lambda=0.5\mu$m to 0.6\arcsec at $\lambda=7.7\mu$m, while remaining nearly constant between 0.5 to 1$\mu$m. Reproducing this plateau requires depletion of the smallest grains. The model with $a_{\rm min} = 0.03\mu$m model predicts a continuous decrease of $d_{\rm neb}$ from 0.5$\mu$m, inconsistent with the observations. We find that $a_{\rm min}\approx 0.3\mu$m can approximately reproduce this plateau. 

However, the reference model still predicts a steeper decline of $d_{\rm neb}$ with wavelength than observed. Adopting $a_{\rm mix}=3\mu$m as in the reference model leads to a $\sim 25\%$ difference in $d_{\rm neb}$ at $7.7\mu$m. Stronger settling $a_{\rm mix}=1\mu$m further enhances this discrepancy, while weaker settling $a_{\rm mix}=10\mu$m does not sufficiently increase the thickness. 
Fixing $a_{\rm min}=0.3\mu$m and $a_{\rm mix}=3\mu$m, we therefore vary the grain size distribution. We find that a shallower power-law index of $p\approx -3.1$ is required to reproduce the $d_{\rm neb}(\lambda)$ trend. However, it predicts a vertically too thick emission at 870$\mu$m, as shown in the Fig.~\ref{fig:mcfost_full}.
These results indicate that dust grains are vertically mixed up to at least $\sim$3$\mu$m and that significant grain growth has occurred in the disk. This conclusion is consistent with previous constraints from \citet{Pontoppidan2007}, who inferred $a_{\rm min}\gtrsim0.5\mu$m and $a_{\rm max}\gtrsim10\mu$m. A similarly shallow wavelength dependence of the dark lane thickness has also been reported for the edge-on disk Tau~042021 \citep{Duchene2024}, where dust grains were found to be vertically mixed up to sizes of $\sim10\mu$m.

\subsubsection{Flux Ratio Reversal}\label{sect: tilt}\begin{figure}
    \centering
    \includegraphics[width=0.9\linewidth]{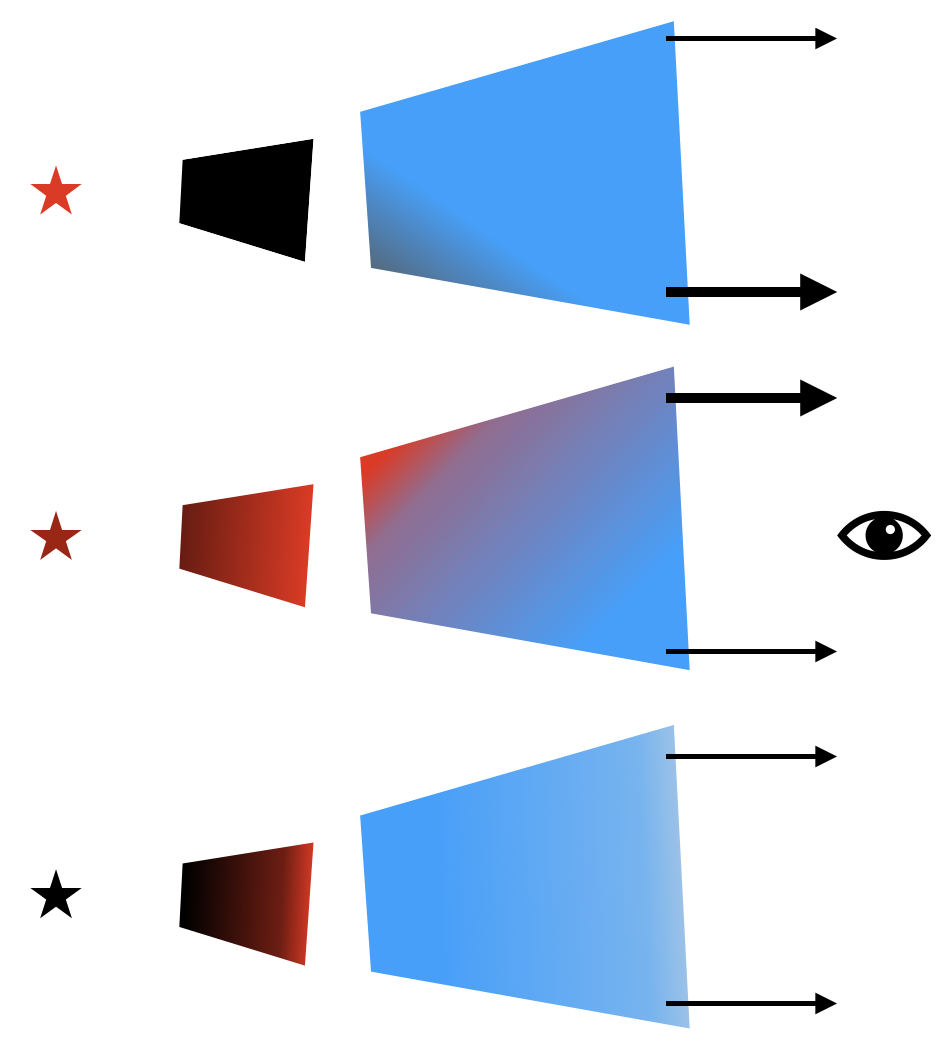}
    \caption{A schematic illustration of the misaligned disk geometry in three wavelength regimes. From top to bottom, the dominant flux source transitions from the stellar scattering ($\lambda<3\mu m$), to a mixed regime including stellar scattering and scattered thermal emission from the inner disk ($\lambda=3-10\mu m$), and to thermal emission from both inner and outer disk($\lambda>10\mu m$). Arrow thickness qualitatively indicates the relative brightness of the two nebulae. }
    \label{fig:schema}
\end{figure}
\begin{figure*}
    \centering
    \includegraphics[width=0.98\linewidth]{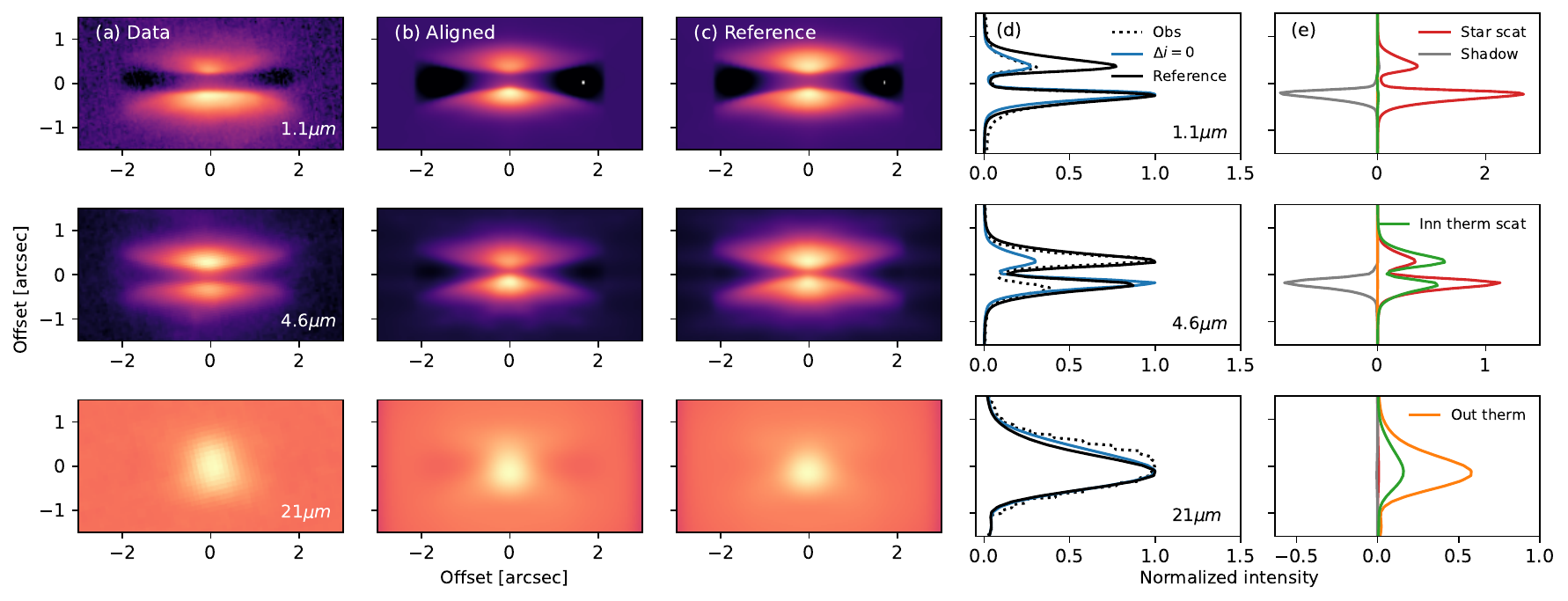}
    \caption{A comparison of observations and model components at 1.1, 4.6, and 21~$\mu$m. Rows correspond to wavelength. Columns (a)-(c) show the observed image, the simulated convolved images of aligned model ($\Delta i=0^\circ$), and the reference model($\Delta i=8^\circ$), respectively. Column (d) shows the corresponding minor-axis brightness profiles, normalized to their peak value at each wavelength. The dotted, blue, and black lines indicate the observations, aligned model, and reference model, respectively. Column (e) decomposes the reference model into scattered stellar light (red), attenuation by the inner disk shadow (grey), scattered thermal emission from the inner disk (green), and thermal emission from the outer disk (orange). }
    \label{fig:component}
\end{figure*}

\begin{figure}
    \centering
    \includegraphics[width=0.96\linewidth]{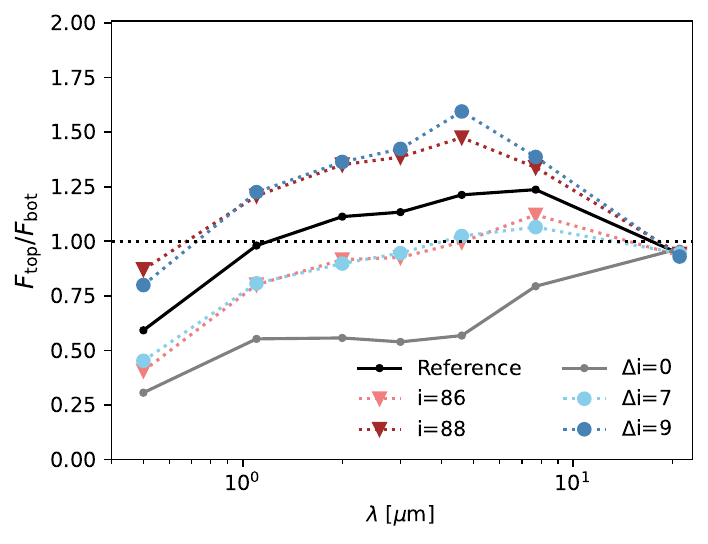}
    \caption{A model flux ratio between the top and bottom nebulae as a function of wavelength. The reference model assumes an outer disk inclination of $i=87^\circ$ and an inner-outer disk misalignment of $\Delta i=8^\circ$. Colored lines show models in which one geometric parameter is varied while the other parameters are kept fixed. The horizontal dotted line marks a flux ratio of unity. }
    \label{fig:flux-ratio-model}
\end{figure}

To clarify the physical origin of the wavelength-dependent flux ratio reversal, Fig.~\ref{fig:schema} provides a schematic overview of the dominant flux sources in different wavelength regimes, and Fig.~\ref{fig:component} compares the observations with the aligned and misaligned model components. As a baseline, we first consider the aligned configuration. In an aligned, axisymmetric disk, the nebula on the side inclined toward the observer is expected to remain brighter because of projection and scattering effects \citep{Burrows1996}. This behavior is seen in the continuous disk model (Fig.~\ref{fig:mcfost_full}, row (a)) and aligned model ($\Delta i=0^\circ$, Fig.~\ref{fig:mcfost_full}, row (b)), and the flux ratios in Fig.~\ref{fig:flux-ratio-model} show that the lower nebula remains brighter up to 21$\mu$m. The aligned configuration therefore cannot reproduce the observed wavelength-dependent brightness reversal. 

Disk misalignment has been proposed as a mechanism to explain mid-infrared brightness reversals \citep{Grosso2003, Villenave2024, Kimmig2025}. In our case, the reversal occurs at significantly shorter wavelengths, motivating a closer examination of how different radiative components contribute to the observed flux ratio. To this end, we compare our reference model, which includes a misaligned inner disk ($\Delta i = 8^\circ$), to an aligned configuration ($\Delta i = 0^\circ$), as shown by the minor-axis brightness profiles in Fig.~\ref{fig:component}(b,c). The minor-axis
brightness profiles are normalized by the peak of the reference model to highlight the impact of misalignment on the relative brightness of the two nebulae, shown in column (d).
To quantify the physical origin of the flux ratio, we further decompose the total flux of the reference model into four components: (1) scattered stellar radiation, (2) flux removed by absorption in the misaligned inner disk, (3) scattered thermal emission from the inner disk, and (4) thermal emission from the outer disk. The relative contributions of these components as a function of wavelength are shown in Fig.~\ref{fig:component}(e). 

At short wavelengths ($\lambda<3\mu m$, Fig.~\ref{fig:schema} top), the observed flux is dominated by scattered stellar light. In a well-aligned system, the side tilted towards the observer (the bottom nebula in our geometry) is always brighter, as the dust scattering is forward-favored.
The flux ratio is primarily controlled by the dust scattering properties and does not change sign. Introducing a misaligned inner disk selectively attenuates stellar photons reaching the near side part, while the far side part remains largely unaffected. As shown in Fig.~\ref{fig:component}(e) shadow component, the brightness of the bottom nebula is reduced, whereas the top nebula brightness remains the same. This shadowing attenuation is more important at shorter wavelengths, which substantially increases the flux ratio.

At longer wavelengths ($\sim$3–10$\mu$m, Fig.~\ref{fig:schema} middle), scattered thermal emission from the inner disk becomes increasingly important. In the reference model, the inner disk emission becomes non-negligible from $3\mu m$ and overweights the stellar flux at $7\mu m$. Fig.~\ref{fig:component}(b, middle right) shows the $\lambda=4.6\mu m$ case, where both stellar light and inner disk emission are important. We note that although the aligned model yields a flux ratio closer to the observed value at shorter wavelengths ($1.1\mu$m), it fails to reproduce the observed brightness reversal at near- to mid-infrared wavelengths. On the contrary, the tilted inner disk in the reference model casts a shadow on the lower nebula, and its thermal emission preferentially illuminates (and enhances scattering) on the top nebula of the outer disk. This asymmetric illumination is the primary driver of the flux ratio increase and the brightness reversal at near- to mid-infrared wavelengths.

At even longer wavelengths ($\lambda>10\mu m$, Fig.~\ref{fig:schema} bottom), the thermal emission from the outer disk becomes important, and the starlight scattering is negligible. The two nebulae are powered by different thermal components: the inner disk thermal emission preferentially enhances the top nebula, whereas the bottom nebula is increasingly dominated by direct thermal emission from the outer disk itself, which favors scattering to the near side, i.e. the bottom nebula. These two contributions partially compensate each other; therefore, the flux ratio gradually drops again. At $21\mu m$, the observed flux is already dominated by scattered light from the outer disk thermal emission (Fig.~\ref{fig:component} (e)).

Within this framework, we explored the model parameter space around the adopted reference configuration. We found that increasing the inner–outer disk misalignment, increasing the outer disk inclination, viewing the system closer to an azimuthal angle of $90^\circ$, increasing the radial extent of the inner disk, or increasing the disk scale height all shift the flux ratio curve upward and move the brightness reversal to shorter wavelengths, while leaving the short-wavelength slope largely unchanged.

Specifically, the dependence of the flux ratio on the disk misalignment and on the viewing inclination is shown in Fig~\ref{fig:flux-ratio-model}. Larger misalignments cast a stronger shadow on the lower nebula and can therefore move the brightness reversal to even shorter wavelengths. Small variations around the reference model parameters reproduce the main observational trends, and there exists degeneracy between misalignment and viewing angle. For example, models with $i=86^\circ$ and $\Delta i=8^\circ$, or $i=87^\circ$ and $\Delta i=7^\circ$, produce comparable wavelength dependence of the flux ratio. Although not explicitly shown here, one could imagine other disk parameters, e.g. inner disk mass and scale height, have similar effects in shadowing the bottom nebula and heating the top nebula of the outer disk. Owing to the degeneracy among geometric parameters, multiple disk configurations can reproduce the overall trend. Therefore, we do not attempt to derive a unique best-fitting model but instead use the reference model as an illustrative configuration.

Variations in the minimum grain size or in the grain size distribution do not qualitatively alter the shape of the flux ratio curve or the wavelength at which the brightness reversal occurs, producing only minor deviations. We also find that strong foreground emission leads to an underestimation of the observed flux ratio. After correcting for this effect, the intrinsic flux ratio reaches $\sim$1.6 at 7.7$\mu$m, while maximum flux ratio occurs near $5\mu m$ remain valid. 

\subsubsection{Silhouette Edge} \label{sect: silhouette_model}
In this section, we focus on the wavelength dependence of the silhouette's edge. 
We applied the same procedure to normalize the major axis brightness profiles of the synthetic observation as described in Section~\ref{sect: silhouette} to ensure quantitatively meaningful comparison with the observations.

\begin{figure}
    \centering
    \includegraphics[width=0.95\linewidth]{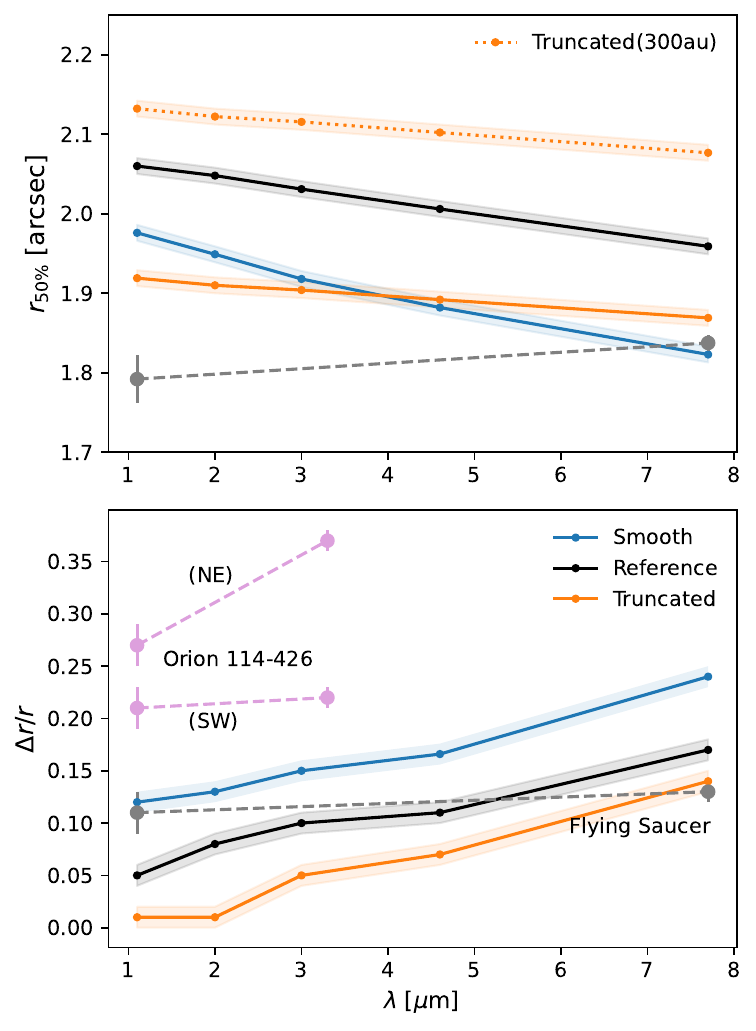}
    \caption{A comparison of the silhouette edge characteristics as a function of wavelength, for $r_{50\%}$ (top), and the normalized silhouette edge width $\Delta r/r$ (bottom).  The solid lines represent model predictions for disks with different radial density profiles at the outer edge: the reference model ($\gamma=-0.5$), a smooth edge ($\gamma=0.5$), and a truncated disk (r=270au). The top panel also includes the $r_{50\%}$ for the truncated disk at $300au$ for comparison with the reference model. Measurements for the Flying Saucer derived from JWST broad-band imaging are overplotted in both panels, while measurements for Orion~114-426 are included in the bottom panel for comparison.}
    \label{fig:edge_compare}
\end{figure}

We first examine the wavelength dependence of the edge location $r_{50\%}$. Observationally, $r_{50\%}$ remains approximately constant from 1.1 to 7.7~$\mu$m within the uncertainties, indicating that the characteristic silhouette edge does not move significantly inward at longer wavelengths. In Fig.~\ref{fig:edge_compare} top panel, we plot the mean $r_{50\%}$ left and right side as listed from Table~\ref{tab:silhouette}. To be conservative, we adopt $0.03\arcsec$ uncertainty at 1.1 $\mu m$, which is the spatial resolution. To explore the role of the radial dust density distribution, we vary the outer disk tapering index $\gamma$, while keeping the dust properties fixed. In addition to the reference model ($\gamma=-0.5$), we consider models with a smoother outer edge ($\gamma=0.5$) and a sharply truncated disk at 270~au, chosen to match the radius at $7.7\mu m$. As shown in Fig.~\ref{fig:edge_compare}, the reference and smooth edge models predict a stronger decrease of $r_{50\%}$ with wavelength. The truncated model shows a weaker wavelength dependence, as expected for a sharper outer boundary, but it cannot simultaneously reproduce the observed edge position at both 1.1 and 7.7~$\mu$m. 

We also measured $r_{95\%}$ as a secondary estimate of the outermost detectable silhouette edge. This quantity also remains approximately constant between 1.1 and 7.7~$\mu$m within the uncertainties. All the models considered in this study predict an essentially wavelength-independent $r_{95\%}$, regardless of the adopted radial density profile and dust size distribution. This suggests that $r_{95\%}$ mainly traces the radius where the disk becomes optically thick, rather than the detailed shape of the outer density fall-off. 

\begin{figure}
    \centering
    \includegraphics[width=\linewidth]{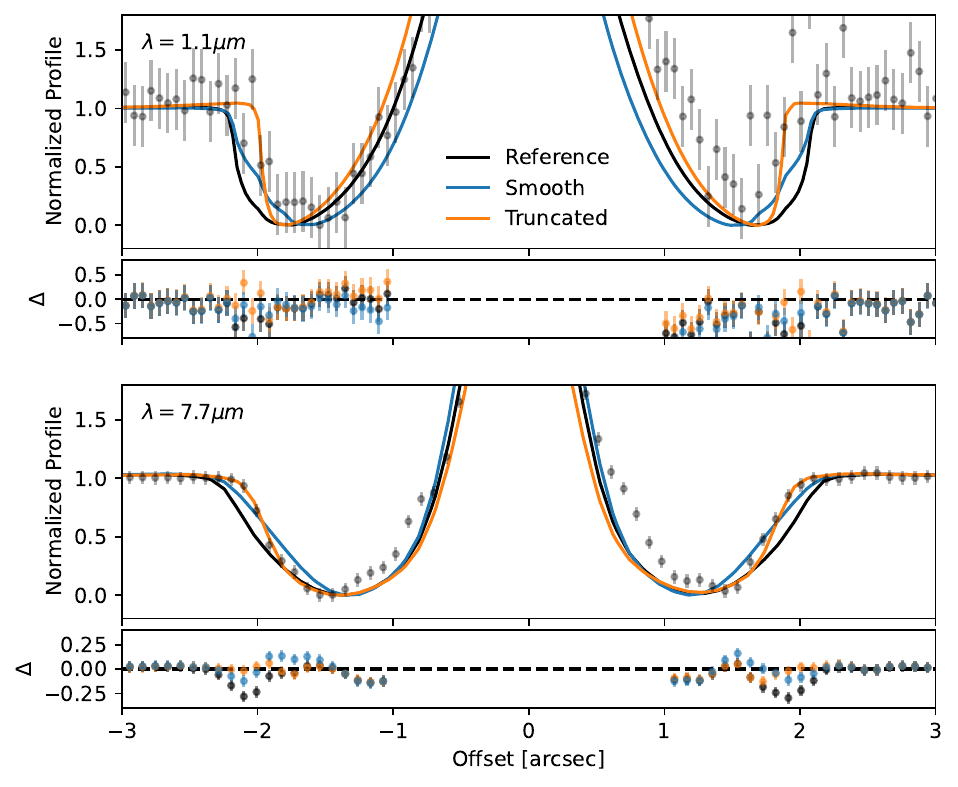}
    \caption{The normalized brightness profile through the major axis at 1.1 and 7.7$\mu$m for models with different radial outer-edge profiles, compared with the observations. Grey points show the observed profiles, while the black, blue, and orange curves show the reference, smooth-edge, and truncated-edge models, respectively. The lower subpanels show the corresponding residuals relative to the observed profiles.}
    \label{fig:sil_edge}
\end{figure}

We then examine the normalized edge width $\Delta r/r$. 
Fig.~\ref{fig:sil_edge} compares the normalized brightness profiles for these models. All the model profiles are shifted by 0.05\arcsec to the left to match the minor left-right asymmetry in observations.  Fig.~\ref{fig:edge_compare} shows the resulting $\Delta r/r(\lambda)$. For $\lambda=2, 3, 4.6\mu m$, the $\Delta r/r$ is estimated by adopting the background emission level $f_{\rm bg}=10^{-2}I_{\rm peak}$ to ensure a prominent absorption feature. For all models with a radially invariant dust population, $\Delta r/r$ increases with wavelength, primarily due to PSF convolution effects. As an example, the reference model predicts a strong wavelength dependence of the normalized edge width, with $\Delta r/r$ increasing from 0.05 to 0.17 from 1.1 to 7.7$\mu$m. However, this trend is not observed in the Flying Saucer, where the silhouette edge remains equally sharp $\Delta r/r=0.11$ across the wavelength range (Table~\ref{tab:silhouette}). We find that the edge width is best match with smooth edge $\gamma=0.5$ at 1.1$\mu$m and truncated edge at 7.7$\mu$m (Fig.~\ref{fig:sil_edge}). This discrepancy indicates that the intrinsic disk edge becomes sharper at longer wavelengths. 

In all models discussed above, the radial dust properties are fixed. We note that $\Delta r/r(\lambda)$ is not uniquely determined by the radial density structure. For a fixed density profile ($\gamma=-0.5$), modifying the dust population can also alter the predicted edge gradients. For example, adopting $a_{\rm min}=0.03\mu$m leads to an sharper increase in $\Delta r/r$ from 0.06 to 0.19 between 1.1 and 7.7$\mu$m, while a shallower size distribution with $p=-3.1$ produces a slightly more moderate increase from 0.05 to 0.15. While the exact slope of $\Delta r/r(\lambda)$ depends on the dust properties, the lack of observed increase due to PSF convolution implies the intrinsic disk edge sharpening.

Finally, our models assume no radial differentiation of the dust population. Observationally, however, the dust distribution is known to vary with radius, as the JWST silhouette observations reveal dust in extinction extending to larger radius than the millimeter continuum \citep{Dartois2025, Guilloteau2025}. Reconciling these observables therefore requires a radially varying dust distribution, which we defer to future work.

\subsubsection{Lining Feature} \label{sect:lining}
We fit the JWST observations at each wavelength independently using Equation~\ref{eq: intensity}. First, the $I_{\rm rt}(x,y)$ images are all normalized by their maximum brightness $I_{\rm peak}$, such that all fitted values are all relative to this value. Then, the background and foreground levels are determined by fitting two scalar intensity offsets: a baseline level measured in an annulus spanning [3.7, 4.0]\arcsec, and a silhouette depth defined as the minimum of the major axis intensity profile. 
At wavelengths where no disk-edge lining is detected, the forward-scattering term is fixed to $f=0$. 
At $\lambda=2.0$ and 3.0$\mu$m where lining is present, we allow the forward-scattering component to vary. The $f$ is fitted by minimizing the residual between [1.2, 3.5]\arcsec, which encloses the lining feature. 
The fitted parameters for the reference model are summarized in Table.~\ref{tab:baseline}. 

At $1.1\mu m$, the fitted foreground term $I_{\rm fg}$ is negative, consistent with foreground extinction along the line of sight. At $7.7\mu m$, the foreground contribution $I_{\rm fg}$ is approximately twice the background level $I_{\rm bg}$, indicating strong PAH emission from the surrounding environment. At $21\mu m$, the emission is dominated by the foreground emission from the fitting result. At all other wavelengths, the fitted background and foreground levels are comparable. The contribution comes from the Ophiuchi cloud.
The forward-scattering lining factor $f$ is robustly constrained at $2.0\mu m$ and $3.0\mu m$, and decreases towards longer wavelengths, as expected for reduction in forward-scattering efficiency toward longer wavelengths. The wavelength dependence of the lining detectability and its physical interpretation are discussed in Section.~\ref{sect: lining_discussion}.

\begin{figure*}
    \centering
    \includegraphics[width=0.8\linewidth]{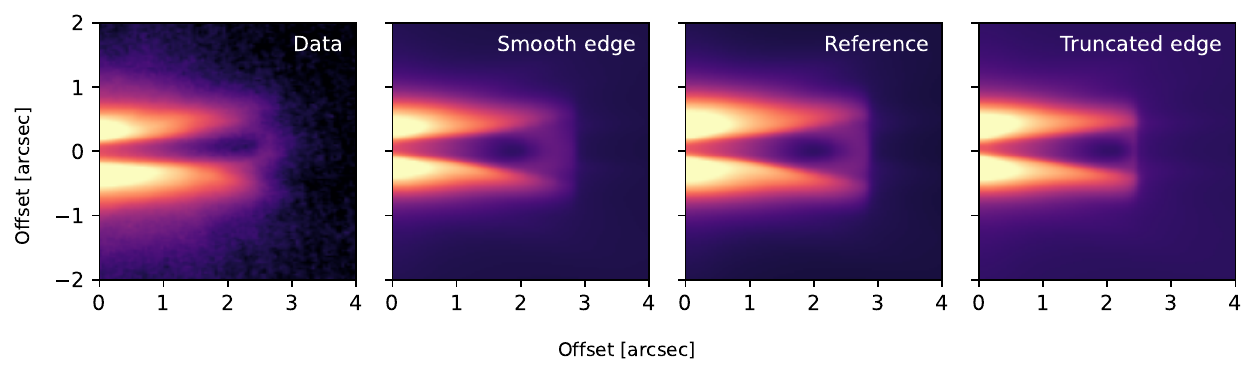}
    \caption{A zoomed-in view of the lining feature at 2~$\mu$m. From left to right, the panels show the data, the smooth-edge model, the reference model, and the truncated-edge model. The simulated images are convolved with the 2~$\mu$m PSF.}
    \label{fig:lining}
\end{figure*}

Fig.~\ref{fig:lining} compares the observed lining feature with models that adopt different radial outer-edge profiles at 2$\mu m$. The reference model reproduces the curved and extended lining morphology that closely matches the observations. The radial location of the lining coincides with the outermost silhouette edge, as both features trace the same disk outer boundary surface. In contrast, models with a sharply truncated outer disk produce a thin and nearly vertical lining, inconsistent with the observed feature. This discrepancy suggests that the Flying Saucer disk is unlikely to be radially truncated for the small dust grains responsible for near-infrared scattering.

\begin{table}[ht]
    \centering
    \caption{Fitted value for the emission from the surrounding environment and lining feature.}
    \label{tab:baseline}
    \begin{tabular}{c c c c}
    \hline
    \hline
    $\lambda$ & $I_{bg}$   &  $I_{fg}$ & $f$ \\
    $(\mu m)$ & ($10^{-3} \ I_{\rm peak}$) & ($10^{-3} \ I_{\rm peak}$) & \\
    \hline
    1.1 & 6.9 & -1.7 & 0 \tablenotemark{a}\\
    2.0 & 2.5 & 0.1 & 7.7 \\
    3.0 & 1.1 & 0.3 & 6.2 \\
    4.6 & 0.9 & 0.9 & 0 \tablenotemark{a}\\
    7.7 & 39.6 & 79.8 & 0 \tablenotemark{a}\\
    21 & 15.7 & 814.8 & 0 \tablenotemark{a} \\
    \hline
    \end{tabular}
    \begin{minipage}{\linewidth}
    \tablecomments{At each wavelength, fitted intensities are normalized to the peak brightness $I_{\rm peak}$ of the corresponding model image $I_{\rm rt}(x,y)$. \tablenotetext{a}{A value fixed to 0 and forward-scattering term is not included in the fit.}}
    \end{minipage}
\end{table}

\section{Discussion} \label{sec:discussion}
\subsection{Origin of the Flux Ratio Variability}
Our modeling demonstrates that a misaligned inner disk can naturally produce a wavelength-dependent flux ratio reversal between the two nebulae, with the brightness switching side at near-infrared wavelengths. In this setup, the key component driving the reversal is the increasing contribution of scattered thermal emission from the inner disk, which preferentially illuminates the top nebula. At longer wavelengths, the contribution from direct thermal emission of the outer disk becomes increasingly important, naturally reducing the flux contrast between the two sides and driving the flux ratio back toward unity by 21$\mu$m.
The reference model qualitatively reproduces the observed trend, including the switching wavelength at near-infrared, although it does not match the flux ratios quantitatively at all wavelengths. 

This behavior is broadly consistent with previous modeling efforts. For example, \citet{Kimmig2025} showed that in warped disk configurations the flux ratio reversal occurs for viewing azimuths $\varphi = 90^\circ \pm 30^\circ$ when the misalignment is large enough($\Delta i >10^\circ$), and that the switching wavelength depends sensitively on the viewing angle and disk geometry. Observationally, similar flux ratio reversals have been reported at longer wavelengths in other edge-on disks, with HH~30 switching between 7.7 and 12.8$\mu$m \citep{Tazaki2025} and IRAS~04302 switching between 12.8 and 21$\mu$m \citep{Villenave2024}. At shorter wavelengths, Orion 114-426 also reports brightness reversal near 2$\mu$m, seen both with HST and JWST \citep{McCaughrean1998, Ballering2025}. These cases suggest that although the wavelength at which the reversal occurs varies from system to system, geometric effects such as inner disk misalignment or warping may be common. 

However, the exceptionally high flux ratio observed at 4.6$\mu$m cannot be reproduced through geometric adjustments alone, as changes that enhance the mid-infrared contrast inevitably overpredict the flux ratio at shorter wavelengths (as shown in Fig.~\ref{fig:flux-ratio-model} for $\Delta i = 9^\circ$ case). Varying geometric parameters does not sharpen flux ratio increase. This suggests that additional wavelength-dependent effects, such as dust composition, may play a role. For example, ice species are observed in the Flying Saucer \citep{Bergner2026, Dartois2025, Martinien2025}. Previous radiative transfer modeling has shown that icy grains affect the disk opacity and scattering properties. Micron- to ten-micron-sized icy grains present at large radii and high vertical altitudes are required to reproduce the observed ice absorption features and multi-wavelength morphology \citep{Dartois2025}. In addition, ice abundance can be derived from the wavelength dependence of the dark-lane thickness \citep{Martinien2025}. However, the impact of ice on the wavelength dependence of the nebular flux ratio has not yet been investigated. Ice is not included in our current model, and quantitative exploration of such compositional effects is deferred to future work \citep{Martinieninprep}. 

While disk misalignment provides a compelling explanation for the wavelength-dependent flux ratio reversal, it is unlikely to be responsible for the temporal variability observed between the HST and JWST epochs. The full disk-integrated flux remains approximately conserved, whereas the flux ratio varies. This behavior cannot be fully explained by purely geometric changes. In particular, increasing the misalignment angle or scale height would only dim the bottom nebula without conserving the total scattered flux. At 1-2$\mu$m, increasing the misalignment from 7$^\circ$ to 8$^\circ$ causes 10\% of flux loss. 

The lateral asymmetry and its variability are naturally introduced in the model by varying the azimuthal orientation $\varphi$ of the inner disk caused by disk precession. However, in this configuration, the resulting asymmetry is primarily confined to the bottom nebula, while the top nebula remains largely symmetric because it is not affected by shadowing. The expected timescales are also likely longer than those inferred from the lightcurve. Therefore, the misaligned geometry in the reference model cannot reproduce the wavelength-dependent $F_{\rm ul}/F_{\rm ur}$ measured in the observation shown in Fig.~\ref{fig: asymmetry}. 

Instead, the variability is more naturally explained by time-dependent illumination effects, such as occulting materials or warps, which modulate the incident radiation field without altering the large-scale disk geometry \citep{Stapelfeldt1999}. This interpretation is also consistent with the system’s optical light curves which show dipper behavior, and with the stronger variability observed at shorter wavelengths, where scattered stellar light dominates. 


Disentangling geometric effects from time-variable illumination will require future observations at short wavelengths (0.5–2$\mu$m) with a cadence shorter than the stellar rotation period ($\lesssim$5~days), ideally combined with simultaneous photometric monitoring. If the nebular asymmetry varies with the stellar light curve, its color can help identify the origin: a chromatic asymmetry would favor stellar spots \citep{Wood1998, Stapelfeldt1999}, while a nearly gray variable asymmetry would favor shadowing by a small-scale inner warp \citep{Venuti2015}. If the asymmetry remains static over the rotation period, it would instead support a larger-scale tilted inner disk. 

\subsection{Origin of Lining}
\label{sect: lining_discussion}
The disk-edge lining is clearly detected for the first time in an edge-on disk. Since it has not been previously identified in other edge-on systems, it is important to understand both why this feature is observed in the Flying Saucer and why it is prominent only at 2 and 3$\mu$m.
In our formalism (Appendix~\ref{app: lining}), this lining term naturally appears as an additional forward-scattering contribution from the background, which enhances the intensity near the silhouette edge. This phenomenon is analogous to the "silver lining" observed in terrestrial clouds, where strong forward scattering of collimated sunlight produces edge brightening along the cloud boundary. In protoplanetary disks, although the illumination field is diffuse, a strongly forward-peaked scattering phase function can redirect background photons into a restricted range of viewing angles, effectively producing a partially collimated illumination. As a result, a similar but weaker edge-brightening feature can emerge along the outer edge $\tau\sim1$.

\begin{figure}
    \centering
    \includegraphics[width=0.9\linewidth]{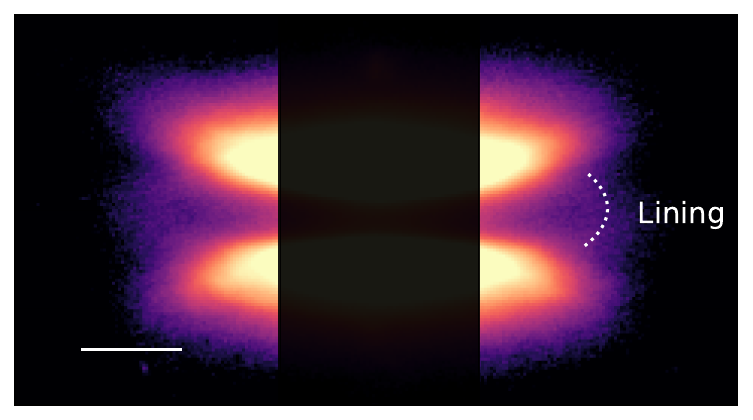}
    \caption{The Tau~042021 disk at F200W, same image as in \citet{Duchene2024} with a different color stretch; and the central bright region within [-1\arcsec, 1\arcsec] is masked to highlight the lining on the edge, indicated by the dotted curved line on the right side. A symmetric lining feature is also seen on the left side. The white bar indicates 1\arcsec. }
    \label{fig:tau042021}
\end{figure}

Several conditions must be satisfied for this lining feature to become observable. First, a sufficiently bright background emission component is necessary as the incident radiation. This requirement likely explains why comparable features are not commonly reported in other edge-on disks. 
Inspection of JWST images of other edge-on disks reveals another case: Tau~042021 also shows a tentative lining feature at 2$\mu$m (See Fig.~\ref{fig:tau042021}. We suspect that such features may also arise in other environments with enhanced external illumination, for example in dense star-forming regions or multiple systems where a luminous companion lies behind the disk.

Second, our interpretation assumes that the scattering phase function remains sufficiently forward-peaked for the quasi-collimated illumination approximation to hold. As the efficiency of forward-scattering decreases toward longer wavelengths, the lining contrast correspondingly weakens. This behavior is consistent with the decline of the fitted forward-scattering factor $f$ with wavelength and explains the nondetection of the lining at longer wavelengths. The observed lining also depends on the geometry and anisotropy of the external illumination field, which may be complex given the location of the Flying Saucer near the edge of a molecular cloud. A quantitative assessment of these environmental effects is beyond the scope of this work.

Third, a high signal-to-noise ratio is essential, as modest forward-scattering contributions can be buried in noise. Although dust is expected to be more forward-scattering at 1.1$\mu$m, the absence of a detectable lining at this wavelength can be partially explained by the noise. Tests with injected noise show that values of $f\lesssim4$ would not produce a detectable lining feature. Earlier HST observations have higher noise, hiding the lining feature. PSF convolution further smears the intrinsically thin lining feature, particularly at long wavelengths. The combined effects of reduced forward-scattering and PSF smoothing likely contribute to the absence of a detectable lining at 7.7$\mu$m despite the presence of strong background emission.

In addition, foreground extinction can suppress the edge contrast at shorter wavelengths by attenuating both the background illumination and the scattered signal. Finally, any departures from the single-scattering regime, such as enhanced multiple scattering when $\omega$ is high, can redistribute the radiation field more isotropically and reduce the prominence of the lining feature. 

We note that line emission from ro-vibrational H$_2$ excited in the disk surface can also produce edge-brightened structures in near-infrared images, as reported by 2.12$\mu$m narrowband imaging in the Orion nebula \citep{Chen1998}. Tentative $H_2$ emission is identified with Spitzer \citep{Pontoppidan2007}. However, our detection is in broadband imaging, which will dilute the line emission. In addition, inspection of the NIRSpec spectra of the Flying Saucer does not reveal a clear lining feature specifically associated with this wavelength. Moreover, strong fluorescent H$_2$ emission typically requires intense UV radiation or shock excitation, conditions that are unlikely in the Flying Saucer given the cool central star ($\sim$3500 K) and the absence of nearby massive stars. 

\subsection{Silhouette Edge as Probes of Disk Evolution}
Early HST surveys of silhouette disks in Orion demonstrated the prevalence of these systems and provided the first constraints on their radii and morphologies \citep{ODell1994, McCaughrean1996}. Individual case studies revealed that reproducing the observed silhouettes requires a steep decrease in density at the outer edge \citep{McCaughrean1996}. However, the Orion 114-426 system appears asymmetric with a stretched tail, suggesting a smoother radial fall-off \citep {Shuping2003, Ballering2025}. To date, no systematic characterization of the outer edge fall-off has been carried out across the silhouette disk population, and it remains unclear whether these disks are best described by sharp truncations, smooth exponential tapers, or a combination of both.

This gap has important implications for disk evolution. In isolation, viscous disk evolution naturally produces smooth exponential taper s in the surface density distribution \citep{Lynden-Bell1974, Hartmann1998}. In contrast, sharp outer edges may indicate the influence of external processes, such as photoevaporation by nearby massive stars or tidal truncation in clustered environments \citep{Bally1998, Adams2004, Winter2018}. Distinguishing between these scenarios requires robust observational constraints on the radial density fall-off, which were difficult to obtain with HST alone due to its limited wavelength coverage, despite its excellent spatial resolution.

With the improved spatial resolution and broad wavelength coverage now available from JWST, silhouette disk edges can be characterized in significantly greater detail. 
As an illustrative comparison, we consider the silhouette disk Orion~114-426, which has recently been observed at multiple wavelengths with JWST \citep{Ballering2025}. Using JWST images taken with F187N and F335M filters, we measure the edge radius and normalized edge width on both sides of the disk at 1.87 and 3.35$\mu$m, and overplot the results in Fig.~\ref{fig:edge_compare}. On the NE side, $\Delta r/r$ increases from 0.27 to 0.37 with wavelength, while on the SW side it remains approximately constant at 0.22. Overall, the silhouette edges of Orion~114-426 are significantly broader than those of the Flying Saucer, implying a shallower radial dust density fall-off. Simple tapered-edge models with $\gamma=0.5$ yet still fall short of reproducing the broad edges observed in Orion~114-426, suggesting even shallower radial gradients. The asymmetric behavior between the two sides further indicates radial and/or grain-size differentiation within the disk, with dust on left side appearing more radially mixed than on the other.
These results highlight the diagnostic power of silhouette edge widths as probes of disk structure and evolution with JWST.

\section{Conclusion} \label{sec: conclusion}
We present a multi-wavelength study of the edge-on protoplanetary disk Flying Saucer based on new JWST broad-band imaging from 1.1$\mu$m to 21$\mu$m, combined with archival HST optical and near-IR data, ALMA millimeter continuum observations, and optical light curves from TESS and K2. These observations are interpreted using radiative transfer modeling to investigate the disk geometry, dust properties, and illumination structure. 

The JWST observations reveal a reversal of the flux ratio between the two disk nebulae at near-infrared (2-3$\mu$m). Radiative transfer modeling shows that this behavior can be explained by a misaligned inner disk, qualitatively reproduces the overall trend and the switching wavelength. The reversal is driven by the contribution of scattered thermal emission from the inner disk. 

For the first time, we detect a curved lining along both disk edges at 2 and 3$\mu$m. This feature is reproduced by strongly forward-scattering near the $\tau\sim 1$ surface. The spatial coincidence of the lining with the silhouette edge indicates that both features trace the same disk outer-edge boundary, offering a good way to probe the density profile at the outer disk edge.  

By comparing the wavelength dependence of the dark-lane thickness measured with HST and JWST, we find that dust grains up to sizes of at least 3$\mu$m are vertically mixed throughout the disk and have undergone significant growth. The observed behavior of the silhouette edge further implies a radial dependence of the dust distribution, with the intrinsic disk edge becoming sharper at longer wavelengths. Reconciling these results with the smaller radial extent observed in millimeter continuum emission will require models with radially varying dust populations.

Multi-epoch imaging reveals variability in the scattered-light morphology at 1–2$\mu$m, while the disk-integrated flux remains approximately conserved. Together with the optical lightcurves, this behavior favors time-dependent illumination effects, such as stellar hot spots or weak inner-disk structures, rather than changes in the large-scale disk geometry. We also measure the proper motion of the system and confirm that the Flying Saucer is comoving with the $\rho$~Ophiuchi cloud.

Overall, this study demonstrates a qualitatively new view of edge-on disks by resolving disk-edge structures and wavelength-dependent brightness asymmetries across a wide wavelength range. By combining these observations with radiative transfer modeling, we show that near-infrared flux ratio reversals, disk-edge linings, and silhouette edges can be interpreted within a unified physical framework.

\begin{acknowledgements}
We thank the anonymous referee for constructive comments that improved the manuscript. J.M. thanks the Swiss National Science Foundation for financial support under grant number P500PT\_222298. This project has received funding from the European Research Council (ERC) under the European Union’s Horizon Europe research and innovation program (grant agreement No.101053020, project Dust2Planets, PI F. Ménard). The research was carried out at the Jet Propulsion Laboratory, California Institute of Technology, under a contract with the National Aeronautics and Space Administration (80NM0018D0004). K.R.S. acknowledges funding support from JWST GO program 4290 provided by NASA through a grant from the Space Telescope Science Institute, which is operated by the Association of Universities for Research in Astronomy, Incorporated, under NASA contractNAS5-26555. The HST and JWST data presented in this article were obtained from the Mikulski Archive for Space Telescopes (MAST) at the Space Telescope Science Institute. The specific observations analyzed can be accessed via \dataset[doi: 10.17909/7j1r-2m95]{https://doi.org/10.17909/7j1r-2m95}.
\facilities{JWST (NIRCam and MIRI), HST (WFC3, WFPC2, NICMOS), ALMA.}
\end{acknowledgements}
\appendix
\section{Model Comparisons}
\label{app:model}
\begin{figure*}
    \centering
    \includegraphics[width=\linewidth]{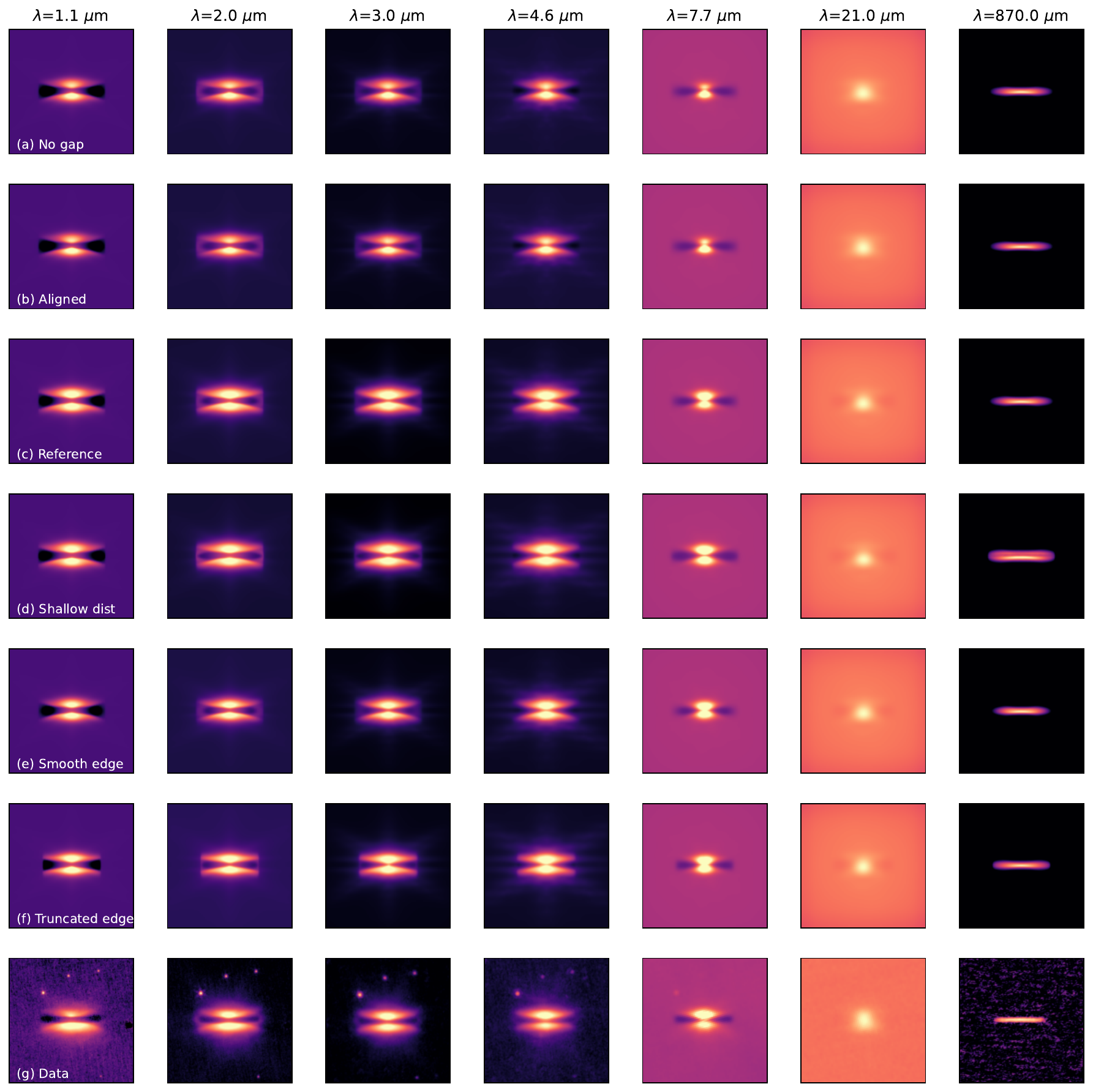}
    \caption{Selected multiwavelength radiative transfer model images shown in logarithmic scale. All the images are convolved with their corresponding PSFs and normalized by their peak flux. Columns show different wavelengths, and rows corresponds to different model configurations: (a) continuous disk, (b) aligned gapped disk, (c) reference disk which is tilted at $8^\circ$, (d) tilted disk with a shallower dust size distribution, (e) tilted disk with smooth outer edge, (f) tilted disk with truncated outer edge, and (g) corresponding JWST and ALMA observations, for comparison. The parameters of the reference model are fully described in Section~\ref{sect: ref_model}. The shallow size distribution model adopts $p=-3.1$ with dust mass adjusted to $3.8\times10^{-3}M_{\odot}$. The truncated disk model has an outer radius of $r=270$~au with dust mass adjusted to $2.0\times10^{-4}M_{\odot}$. All panels display a field of view \textbf{$8\arcsec \times 8\arcsec$.} }
    \label{fig:mcfost_full}
\end{figure*}

This section presents simulated multi-wavelength images for selected models (see Figure~\ref{fig:mcfost_full}). All the simulated images are convolved with their corresponding PSFs as in the main text. All panels are shown in $8\arcsec\times 8\arcsec$ field of view. 
The first three rows illustrate the effects of introducing a gap and an inner-outer disk misalignment. To isolate the role of the gap, we compare a continuous disk model (row (a)) with an aligned gapped model (row (b)). The adopted gap, located at 3-5 au, has a negligible effect on the scattered-light morphology and on the wavelength dependence of the nebular flux ratio. In both aligned models, the lower nebula remains brighter at all wavelengths because it corresponds to the near side in our viewing geometry. The tilted reference model (row (c)) instead qualitatively reproduces the flux ratio reversal, although the contrast is weaker than the data. The models starting from row (c) include the misalignment.

Rows (c), (e), and (f) show the effect of the radial fall-off at the outer disk edge. These models are discussed in the lining analysis in Sect.~\ref{sect:lining} and in the silhouette analysis in Sect.~\ref{sect: silhouette_model}. The model (f) is truncated at 270~au to match the $7.7\mu m$ observation and the disk mass is adjusted to $2\times 10^{-4}M_\odot$. Their simulated images are qualitatively similar at most wavelengths. We note that, the disk misalignment is slightly adjusted among these models, with $\Delta i = 8^\circ, 7^\circ,$ and $6^\circ$ for rows (c), (e), and (f) respectively, so that the peak brightness reversal occurs near $2\sim3 \mu m$. 

Row (d) shows a model with a shallower surface-density distribution, $p=3.1$. The disk mass is adjusted to $M=3.8\times 10^{-3}M_\odot$, while the other parameters remain the same as the reference model. The shallower dust size distribution provides the best match to the measured dark-lane thickness, but fails to reproduce the ALMA continuum morphology. This discrepancy may indicate that stronger dust settling is required. 

\section{Radiative Transfer Approximation for Lining Feature}
\label{app: lining}
To explain why a forward-scattering dust layer can produce a bright lining feature (like the silver lining of clouds), we use the radiative transfer equations. 
Imagine the disk as a plane-parallel slab; we assume the direction of incident light $\mu=\cos\theta$ and emergent light $\mu' = \cos\theta'$. This slab is illuminated only from one side, with optical depth $\tau$.  

The solution of intensity emerging at optical $\tau$ in direction $\mu'$ is:
\begin{equation}
    I(\tau, \mu') = I(0, \mu')e^{-\tau/\mu'} + \int_{0}^{\tau} S(t, \mu') e^{-(\tau-t)/\mu'} dt/\mu'
\end{equation}
The first term describes the direct transmitted part, and the second term is the scattering added along the path. Since the source comes from background nebula emission, it arrives from all incoming direction $\mu$. Therefore, $S(t, \mu') = \int P(\mu, \mu')I(t, \mu) d\mu$, where $P(\mu, \mu')$ is the effective scattering kernel, describing the angular redistribution from $\mu$ to $\mu'$. In this reduced 1D formulation, the integration over solid angle and the scattering albedo are absorbed into $P$. And before scattering at depth $t$, the incoming radiation has already been attenuated as $I(t, \mu) = I(0, \mu)e^{-t/\mu}$. Substitution makes: 
\begin{equation}
    I(\tau, \mu') = I(0, \mu')e^{-\tau/\mu'} + \int_{0}^{\tau} \int P(\mu, \mu')I(0, \mu)e^{-t/\mu} d\mu \   e^{-(\tau-t)/\mu'} dt/\mu'
\end{equation}

In our case, the emerging light is along the observer's line of sight, therefore $\mu'=1$. We also assume a homogeneous diffuse background $I(0, \mu) = I(0)$, which leads to:
\begin{equation}
    I(\tau) = I(0)e^{-\tau} + I(0)\int_{0}^{\tau} \int P(\mu, \mu'=1)e^{-t/\mu} d\mu \  e^{-(\tau-t)} dt
\end{equation}

We impose the assumption of strong forward scattering, so that redistribution is concentrated around $\mu \approx \mu'$ within a small effective angular width, and $\int P(\mu, \mu'=1)e^{-t/\mu} d\mu \approx \int P(\mu \approx \mu') e^{-t/\mu'} d\mu \propto e^{-t}$. We approximate the ratio $\int P(\mu, \mu'=1)e^{-t/\mu} d\mu / e^{-t} $ by an effective constant coefficient $f$, describing the averaged strength of the forward-scattering. Then we obtain: 
\begin{equation}
    I(\tau) \approx  I(0)e^{-\tau} + fI(0)\int_{0}^{\tau} e^{-t} \  e^{-(\tau-t)} dt
\end{equation}
This simplifies to:
\begin{equation}
    I(\tau) \approx  I(0)e^{-\tau} + fI(0) \tau e^{-\tau}
\end{equation}
which is exactly the extinction and scattering terms that we added in equation~\ref{eq: intensity}. 
This lining feature can be seen as an analog to the silver lining observed in terrestrial clouds. In clouds, the background illumination is provided by collimated sunlight, leading to a pronounced edge-brightening effect. In protoplanetary disk, this feature is weaker because the radiation field is more diffuse. However, strong forward scattering effectively redirects background photons into a narrow range of angles toward the observer, partially mimicking a collimated illumination. As a result, a similar but weaker edge-brightening feature is produced along the disk rim.


\bibliography{biblio}{}
\bibliographystyle{aasjournalv7}



\end{document}